\documentclass[]{elsart}
\usepackage{graphicx,amsmath,amssymb}
\usepackage{natbib}
\newcommand{\scaption}[1]{\caption{\small #1}}

\renewcommand{\cite}{\citep}
\newcommand {\bm}[1]  {\mbox{\boldmath ${#1}$}}

\mathchardef\itGamma="7100
\mathchardef\itDelta="7101
\mathchardef\itTheta="7102
\mathchardef\itLambda="7103
\mathchardef\itXi="7104
\mathchardef\itPi="7105
\mathchardef\itSigma="7106
\mathchardef\itUpsilon="7107
\mathchardef\itPhi="7108
\mathchardef\itPsi="7109
\mathchardef\itOmega="710A

\begin{document}
\begin{frontmatter}

\title{Periodic motion representing isotropic turbulence}
\author{Lennaert van Veen\corauthref{cor}\corauthref{mech}},
\author{Shigeo Kida\corauthref{mech}}
\author{and}
\author{Genta Kawahara\corauthref{aer}}
\corauth[cor]{Email address: veen@mech.kyoto-u.ac.jp}
\address[mech]{
Department of Mechanical Engineering, Kyoto University, \\
Sakyo-ku Yoshida-Honmachi, Kyoto 606-8501, Japan
}
\address[aer]{
Department of Auronautics and Astronautics, Kyoto University, \\
Sakyo-ku Yoshida-Honmachi, Kyoto 606-8501, Japan
}
\begin{abstract}
\hspace*{0.3cm}
Temporally periodic solutions 
are extracted numerically from forced box turbulence with high symmetry. 
Since they are unstable to small perturbations, 
they are not found by forward integration but can be captured by 
Newton-Raphson iterations. 
Several periodic flows of various periods are identified 
for the micro-scale Reynolds number $R_{\lambda}$ between $50$ and $67$.   
The statistical properties of these periodic flows are compared with 
those of turbulent flow. 
It is found that the one with the longest period, 
which is two to three times the large-eddy-turnover time of turbulence,
exhibits the same behaviour quantitatively as turbulent flow.
In particular, we compare
the energy spectrum, the Reynolds number dependence of 
the energy-dissipation rate, 
the pattern of the energy-cascade process, and 
the magnitude of the largest Lyapunov exponent. 
This periodic motion consists of high-activity and 
low-activity periods, which turbulence approaches, more often around 
its low-activity part, 
at the rate of once over a few eddy-turnover times. 
With reference to this periodic motion 
the Kaplan-York dimension and the Kolmogorov-Sinai entropy of 
the turbulence with high symmetry 
are estimated at $R_{\lambda}=67$ to be $19.7$ and $0.992$ 
respectively. 
The significance of such periodic solutions, embedded 
in turbulence, for turbulence analysis is discussed.
\end{abstract}
\end{frontmatter}

\section{Introduction}

\hspace*{0.3cm}
Turbulence is a complex state of fluid motion.
The flow field varies randomly both in space 
and in time. 
An individual flow field is too complicated to extract any simple and 
useful information from, and does not exhibit 
any universal laws.
The mean flow fields, on the other hand, obtained by spatial, 
temporal or ensemble averaging, exhibit simpler 
behaviour and allow for the extraction of universal statistical 
properties. 
Useful information, if any, is expected to be seen 
more clearly in the mean flow. 
In fact, the celebrated statistical laws of turbulence, such as 
the Kolmogorov universal law for the 
energy spectrum at small scales (see \citet{MoYa75}) and the 
logarithmic velocity profile in wall turbulence 
(see \citet{Schl79}) 
were confirmed experimentally by ensemble averages of many 
measured data. 

\hspace*{0.3cm}
In contrast to the statistical ones, the dynamical properties of 
turbulence are blurred in the mean field and must be 
analysed in the instantaneous flow. 
The fact that the fluid motion is chaotic and never repeats, 
however, 
makes it extremely difficult to extract any universal dynamical properties. 
There is no way to pick up, with confidence, 
any representative parts of turbulent flows from a finite series of temporal 
evolution. 
Thus, it would be nice if there are some reproducible flows, or skeletons 
of turbulence, which represent the turbulent state well. 
This is reminiscent of unstable periodic orbits in chaotic dynamical 
systems. 
The chaotic attractor contains infinitely many unstable periodic orbits. 
Some statistical properties associated with a strange attractor 
are described in terms of the unstable periodic orbits embedded in it.
Rigorous results are provided by the {\em cycle expansion} theory 
\cite{artuso}.
However, these results only seem to apply to a certain class of dynamical 
systems with a chaotic attractor of a dimension less than three, 
a far cry from developed turbulence. 
The dimension of the attractor of turbulent flow is expected 
to grow with the number of modes in the inertial range. For turbulent 
Poiseuille flow, for instance, the attractor dimension has been estimated 
to be $\mbox{O}(100)$ at a wall-unit Reynolds number of $R_{\tau}=80$ \cite{keefe}.

\hspace*{0.3cm}
In such high-dimensional chaos it is unknown whether an infinite number of 
periodic orbits is 
necessary to describe the statistical properties of the strange attractor 
or a finite number of them is sufficient. 
In this respect, two key papers have recently been published. 
\citet{kawahara} found two periodic solutions in the plane Couette 
system with $15,422$ degrees of freedom and showed that they represent 
the quiescent and turbulent phases of the flow. 
The latter periodic solution represents the generation cycle of 
turbulent activity, i.e. the repetition of alternate generation and 
breakdown of streamwise vortices and low-speed streaks. 
Moreover, the phenomenon of bursting is explained as the state point 
wandering back and forth between these solutions. 
This provides us with the first example that shows that only a single
periodic motion represents the properties of the turbulent state well. 
The second example is the discovery of a periodic solution in shell 
model turbulence with $24$ degrees of freedom \cite{KaYa03}.
A one parameter family of the solution exhibits 
the scaling exponents of the structure function of the velocity field 
similar to real turbulence.

\hspace*{0.3cm}
Inspired by these discoveries of periodic solutions which represent 
the turbulent state by themselves, we were led to the present 
search of 
periodic motion in isotropic turbulence, hoping to find one 
which reproduces turbulent statistics such as the Kolmogorov 
energy spectrum in the universal range. 
Such periodic orbits are asymptotically unstable and are not found by
simple forward integration. They can only be captured by Newton-Raphson
iterations or similar methods.
Here, we encounter a hard practical problem, namely that
the computation time required for the perfomance of Newton-Raphson 
iterations increases rapidly as the square of the number of degrees of 
freedom which is enormous in a simulation of the turbulent state. 
The present our computer resources limit 
the available number of degrees of freedom to $\rm{O}(10^4)$. 

\hspace*{0.3cm}
In the next section, we impose the high symmetry to the flow 
to reduce the number of degrees of freedom in simulations \cite{kida1}. 
The onset of developed turbulence at micro-scale Reynolds number 
$R_{\lambda}=67$, described in section \ref{sec:turbulent}, can then be 
resolved by taking account of about $10^4$ degrees of freedom.
The localisation of periodic solutions in such large sets of equations 
is a hard task indeed. 
In section \ref{sec:periodic}, we take the approach 
of regarding periodic solutions as fixed points of a Poincar\'e map. 
Newton-Raphson iterations can then be used to find such fixed points. 
The iterations, however, converge only if a good initial 
guess is provided. 
We filter initial data from a turbulent time series by looking for 
approximately periodic time segments. 
This works well at fairly low $R_{\lambda}$, 
where the flow is only weakly turbulent. Subsequently we use arc-length 
continuation to track the periodic solutions into the regime of 
developed turbulence. 
We present several periodic solutions of different period and compare 
them to the turbulent state in a range of $R_{\lambda}$. 
Then we show in section \ref{sec:embedded} 
that the solution of longest period considered 
here, about two to three times the large-eddy-turnover time, 
represents the turbulence remarkably well. 
In particular we compare the time-averaged energy-dissipation rate, 
the energy spectrum and the largest Lyapunov exponent. 
Further, we examine the dynamical properties of this particular 
periodic motion and 
show that it exhibits the energy-cascade process by itself.
It consists of a low-active period and a high-active period, 
and the turbulent state approaches it selectively in the low-active part 
at the rate of once over several eddy-turnover times. 
We compute a part of the Lyapunov spectrum of the periodic motion and 
the corresponding Kaplan-Yorke dimension and Kolmogorov-Sinai entropy.
These values can be considered as an approximation of the values found in 
isotropic turbulence under high-symmetry conditions. The local Lyapunov
exponents are shown to have systematic correlations to the energy input
rate and dissipation rate of the periodic motion, which leads to the
conjecture that the ordering of the Lyapunov vectors by the magnitude of the
corresponding exponents corresponds to an ordering of spatial scales of the
perturbation fields they describe. 
Finally, future perspectives of the turbulence research on the basis of the 
unstable periodic motion will be discussed 
in section \ref{sec:conclusion}. 

\section{High-Symmetric Flow}
\label{sec:highsymm}
\hspace*{0.3cm}
We consider the motion of an incompressible viscous fluid 
in a periodic box given by $0<x_1,x_2,x_3\leq 2\pi$. 
The velocity field $\bm{u}(\bm{x},t)$ and the vorticity field 
$\bm{\omega}(\bm{x},t)=\nabla\times\bm{u}(\bm{x},t)$ are expanded in 
the Fourier series of $N^3$ terms as 
\begin{eqnarray}
\bm{u}(\bm{x},t) &=& \mbox{i} 
\sum_{\bm{\scriptstyle k}}\widetilde{\bm{u}}(\bm{k},t)
\mbox{e}^{{\rm i} \bm{\scriptstyle k}\cdot\bm{\scriptstyle x}}, \\
\bm{\omega}(\bm{x},t) &=& \sum_{\bm{\scriptstyle k}} \widetilde{\bm{\omega}}
(\bm{k},t)\mbox{e}^{{\rm i}\bm{\scriptstyle k}\cdot\bm{\scriptstyle x}},
\end{eqnarray}
where $\bm{k} =(k_1,k_2,k_3)$ is the wavenumber 
and the summations are taken over all triples of integers satisfying 
$-\frac{1}{2}N<k_1, k_2, k_3 \le \frac{1}{2}N$. 
Then, the Navier-Stokes and the continuity equations are 
respectively written as
\begin{eqnarray}
\frac{\mbox{d}}{\mbox{dt}}\widetilde{\omega}_{i}(\bm{k},t) 
&=& \epsilon_{ijk}k_{j}k_{l} \,\widetilde{u_{k}u_{l}}(\bm{k},t)
 -\nu k^2 \widetilde{\omega}_{i}(\bm{k},t), \label{NS}\\
[0.5cm]
k_{i}\widetilde{u}_{i}(\bm{k},t)&=& 0,
\label{cont}
\end{eqnarray}
where $\nu$ is the kinematic viscosity, $\epsilon_{ijk}$ is the 
unit anti-symmetric tensor, 
and the tilde denotes the Fourier transform. 
The summation convention is assumed for the repeated subscripts. 
By definition, the Fourier transforms of the velocity and vorticity 
fields are related by 
\begin{equation}
\widetilde{\omega}_{i}(\bm{k},t)=-\epsilon_{ijk}k_{j}
\widetilde{u}_{k}(\bm{k},t).
\end{equation}

\hspace*{0.3cm}
In order to reduce the number of degrees of freedom 
we impose the high symmetry on the flow field \cite{kida1}, in which 
the Fourier components of vorticity are real, and satisfy
\begin{equation}
\widetilde{\omega}_1(k_1,k_2,k_3;t)=
 \begin{cases}
   +\widetilde{\omega}_2(k_3,k_1,k_2;t),\\
   +\widetilde{\omega}_3(k_2,k_3,k_1;t),\\
   +\widetilde{\omega}_1(-k_1,k_2,k_3;t),\\
   -\widetilde{\omega}_1(k_1,-k_2,k_3;t),\\
   -\widetilde{\omega}_1(k_1,k_2,-k_3;t),\\
   +\widetilde{\omega}_1(k_1,k_3,k_2;t) 
       &\hbox{(if $k_1$, $k_2$, $k_3$ are all even),}\\
   -\widetilde{\omega}_1(k_1,k_3,k_2;t) 
       &\hbox{(if $k_1$, $k_2$, $k_3$ are all odd),}\\
   0 & \hbox{(unless $k_1$, $k_2$, $k_3$ are all even or odd).}
 \end{cases}
\label{hsymmetry}
\end{equation}

Under these conditions, only a single component of the vorticity field has to be computed in a 
volume fraction $1/64$ of the periodic domain 
and the number of degrees of freedom is reduced by a factor of $192$. 

\hspace*{0.3cm}
The flow is maintained by fixing the magnitude of the smallest wavenumber 
components of velocity, which otherwise tends to decay in time 
due to transfer of energy to larger wavenumbers leading to ultimate  
dissipation by viscosity. 
The magnitude of the smallest wavenumbers of a nonzero velocity component 
under high-symmetry condition is 
$k_{f}=\sqrt{11}$, and the magnitude of the velocity of 
the fixed components is set to be 
\begin{equation}
|\widetilde{u}_{i}(\bm{k},t)|=\textstyle{\frac{1}{8}}\qquad
\hbox{($i=1,2,3$)\hspace*{1cm} for $|\bm{k}|=k_f$.}
\label{eq:forcing}
\end{equation}
Since the magnitude of these components of velocity decreases, 
in average, in each 
time step of numerical simulation, this manipulation results in 
energy supplies to the system. 
As will be discussed in subsection \ref{subsec:structure}, 
the energy-input rate,
\begin{equation}
e(t)=\sum_{|\bm{\scriptstyle k}|=k_{f}} \widetilde{u}_{i}(\bm{k},t)
\frac{\mbox{d}}{\mbox{dt}}{\widetilde{u}}_{i}(\bm{k},t),
\end{equation}
changes in time depending on the state of flow. 

\hspace*{0.3cm}
Equations (\ref{NS}) and (\ref{cont}) are solved numerically starting 
with some appropriate initial condition. 
The nonlinear terms are evaluated by the spectral method in which 
the aliasing interaction is suppressed by eliminating all the 
Fourier components beyond the cut-off wavenumber $k_{\rm max}=[N/3]$, 
the maximum integer not exceeding $N/3$. 
In the following, we fix $N=128$ so that the number $n$ of degrees of freedom 
of the present flow is about $10^4$. 
The fourth-order Runge-Kutta-Gill scheme with step size $\Delta t=0.005$ 
is employed for time stepping. 

\hspace*{0.3cm} 
For later use, we introduce several global quantities which 
characterise the flow properties, namely, the total kinetic energy of 
fluid motion, 
\begin{eqnarray}
{\mathcal E}(t) &=& \frac{1}{(2\pi)^3}\int\frac{1}{2}
|\bm{u}(\bm{x},t)|^2{\rm d}\bm{x} 
= \frac{1}{2}\sum_{\bm{\scriptstyle k}}
|\widetilde{\bm{u}}(\bm{k},t)|^2,
\end{eqnarray}
the enstrophy, 
\begin{eqnarray}
{\mathcal Q}(t) &=& \frac{1}{(2\pi)^3}\int\frac{1}{2}
|\bm{\omega}(\bm{x},t)|^2{\rm d}\bm{x} 
= \frac{1}{2}\sum_{\bm{\scriptstyle k}}
|\widetilde{\bm{\omega}}(\bm{k},t)|^2,
\label{enstrdef}
\end{eqnarray}
the energy-dissipation rate, 
\begin{equation}
\epsilon(t) = 2\nu {\mathcal Q}(t), 
\end{equation}
and the Taylor micro-scale Reynolds number,
\begin{equation}
R_{\lambda}(t) = \sqrt{\frac{10}{3}}\frac{1}{\nu}
\frac{{\mathcal E}(t)}{\sqrt{{\mathcal Q}(t)}}, 
\end{equation}
where the integration is carried out over the whole periodic box. 
In the following,
time-averaged quantities are denoted by an over bar.

\section{Turbulent State}
\label{sec:turbulent}

\hspace*{0.3cm}
The reduction by symmetry introduced above makes
it possible to describe the laminar-turbulent transition and the 
statistics of fully developed
turbulence in terms of relatively few degrees of freedom. 
With the forcing as described by Eq. (\ref{eq:forcing}), 
the following scenario is observed for decreasing viscosity \cite{kida2,veen}.

\hspace*{0.3cm} 
The flow is steady at large viscosity $\nu$,
or small micro-scale Reynolds number $R_{\lambda}$, 
and remains so down to $\nu\approx 0.01$ 
($R_{\lambda}\approx 25$), where 
a Hopf bifurcation takes place and the flow becomes periodic with 
a period of about $2.2$. 
This period is identical to the Poincar\'e return time 
$T_{\rm{\scriptscriptstyle R}}$ which will be introduced in section 
\ref{sec:periodic}. 
The stable periodic 
motion subsequently undergoes a torus bifurcation and the motion becomes 
quasi-periodic. 
In a range of viscosity, $0.008>\nu >0.005$ 
($35<R_{\lambda}<50$), we observe the breakdown and creation of invariant 
tori, and the behaviour alternates between 
quasi-periodic and chaotic. 
In this chaotic regime the spatial structure 
of the flow remains simple so that we can speak of `weak turbulence'. 
Around $\nu=0.005$ the 
flow becomes chaotic through the Ruelle-Takens scenario, and for lower 
viscosity ($\nu<0.005$) only disordered behaviour is found.
Then, for $\nu<0.004$ ($R_{\lambda}>60$), the time-averaged 
energy-dissipation rate $\overline{\epsilon}$ hardly 
changes as a function of viscosity and fully developed turbulence sets 
in.  

\begin{figure}[h]
\begin{center}
\includegraphics[width=.8\textwidth]{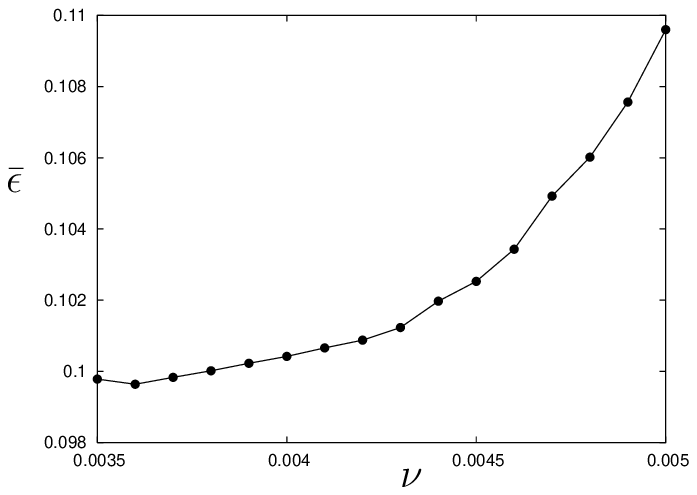}
\end{center}
\scaption{The time-averaged energy-dissipation rate $\overline{\epsilon}$ 
against viscosity $\nu$ in the turbulent state. 
As $\nu$ decreases, $\overline{\epsilon}$ seems to saturate around $0.1$.}
\label{turbulent1}
\end{figure}

\hspace*{0.3cm}
In Fig. \ref{turbulent1}, we display $\overline{\epsilon}$ against $\nu$ 
over the range $0.0035<\nu <0.005$, where a transition from
weak to fully developed turbulence takes place. 
Observe that $\overline{\epsilon}$ seems to saturate around $0.1$
at smaller viscosity. 
This property will play a key role 
in identifying periodic motion that represents the 
turbulent state in section \ref{sec:periodic}.
As is common to many kinds of turbulence, quite large fluctuations 
are observed in time series of $\epsilon(t)$. 
The standard deviation $\sigma_{\epsilon}$ 
is about $10\sim 20$\% of the mean value $\overline{\epsilon}$ 
in the present flow. 
For example, $\sigma_{\epsilon}$ takes the value 0.009 at $\nu=0.0045$ 
and 0.016 at $\nu=0.0035$, too large to be drawn in the figure. 
For a plot of $R_{\lambda}$ against $\nu$, see \citet{kida2}.

\begin{figure}[h]
\begin{center}
\includegraphics[width=.8\textwidth]{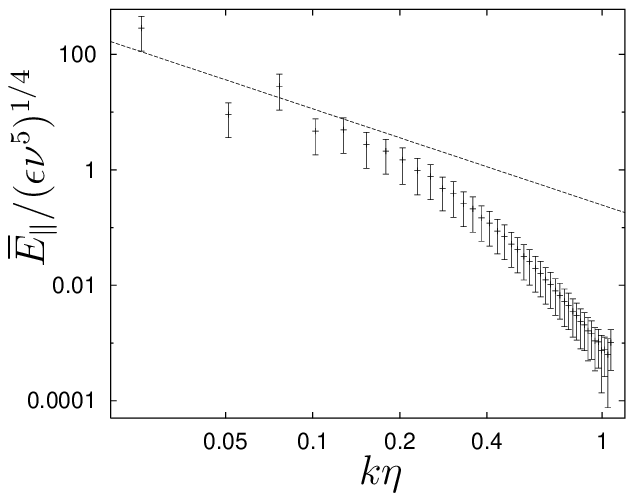}
\end{center}
\scaption{One-dimensional longitudinal energy spectrum for the turbulent 
state at $R_{\lambda}=67$ ($\nu=0.0035$) with
error bars denoting standard deviation. The straight
line denotes the Kolmogorov $-5/3$ power law with Kolmogorov 
constant of 1.4. 
The both axes are normalised by the Kolmogorov characteristic scales.}
\label{turbulent2}
\end{figure}

\hspace*{0.3cm}
The energy spectrum, which represents the scale distribution of 
turbulent activity, is one of the most fundamental statistical 
quantities characterising turbulence. 
Since the longitudinal velocity correlation is relatively easy to be 
measured in experiments, the one-dimensional longitudinal 
energy spectrum is frequently compared between different kinds of
turbulence. 
In the high-symmetric flow, it is calculated by 
\begin{equation}
E_{\parallel}(k,t)=\frac{1}{2}\sum_{k_2,k_3}|\widetilde{u}_1(k,k_2,k_3;t)|^2. 
\label{eq:longitudinal}
\end{equation}
In Fig. \ref{turbulent2}, we plot the time-averaged one-dimensional 
longitudinal energy spectrum $\overline{E}_{\parallel}(k)$ at the maximal 
micro-scale Reynolds number $R_{\lambda}=67$ attained in our numerical 
experiments.
The straight line indicates the Kolmogorov $-5/3$ power law with Komogorov 
constant of $1.4$. 
The inertial range appears only marginally at this Reynolds number. 

\hspace*{0.3cm}
In order to go to larger Reynolds numbers, we need to increase the 
truncation level to maintain $k_{\rm max}\eta\approx 1$, 
where $\eta=({\nu^3/\overline{\epsilon}})^{1/4}$ is the
Kolmogorov length. 
The main impedediment for increasing
the truncation level is the computation time and memory requirement
of the continuation of periodic orbits, as will be described in section 
\ref{sec:periodic}. 
In previous work by \citet{kida4} it was shown that the high-symmetric 
flow reproduces the Kolmogorov spectra accurately at large Reynolds 
numbers ($R_{\lambda}\sim 100$). 
The intermittency effects were investigated by \citet{kida5} and 
\citet{boratav}.

\begin{figure}[h]
\begin{center}
\includegraphics[width=.8\textwidth]{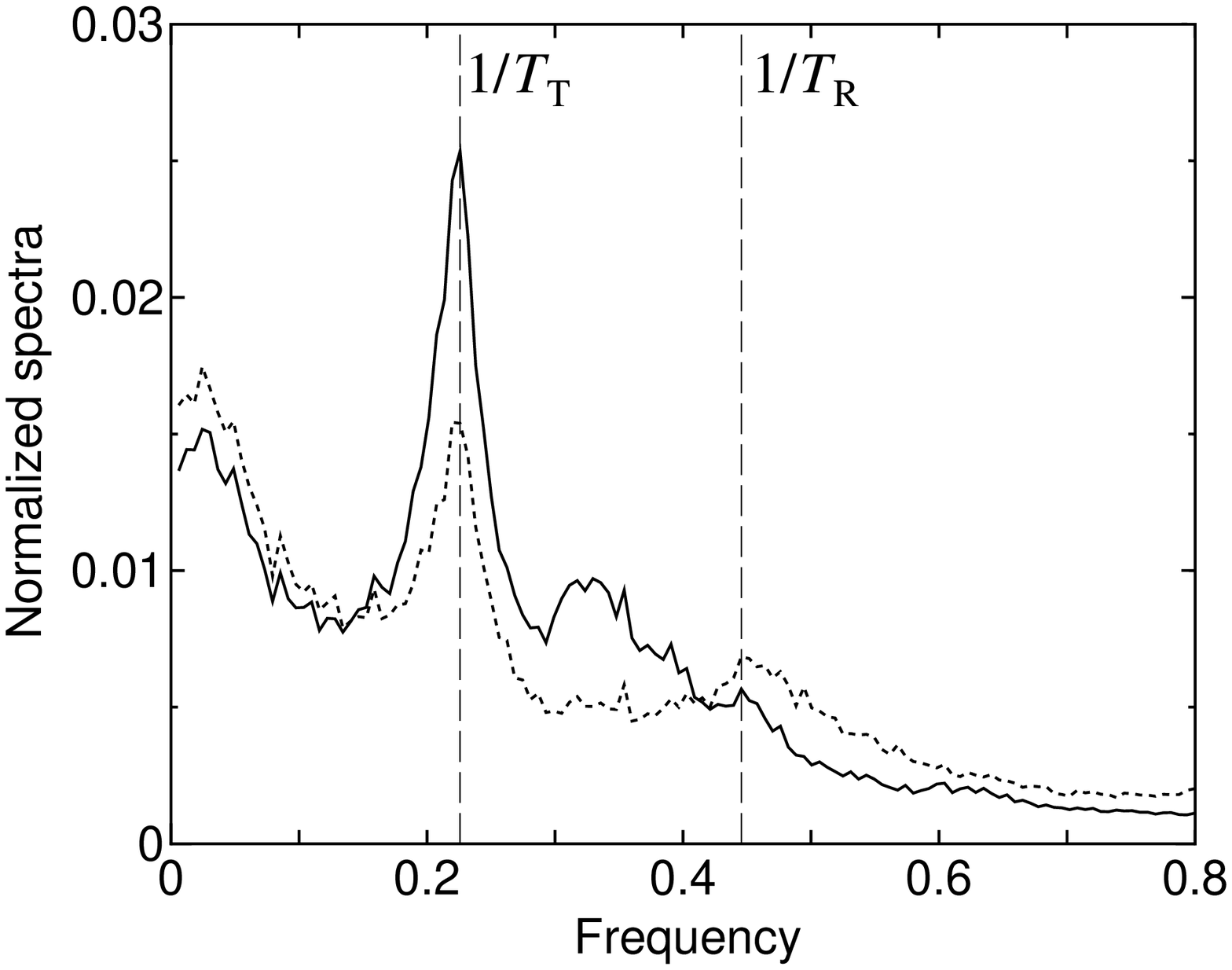}
\end{center}
\scaption{Frequency spectra of energy (solid line) and enstrophy 
(dotted line) at $\nu=0.0035$.
Dashed lines 
are drawn at the peaks corresponding to 
the large-eddy-turnover time $T_{\rm{\scriptscriptstyle T}}$
and the most probable return time $T_{\rm{\scriptscriptstyle R}}$ 
of the Poincar\'e map, 
described in section \ref{sec:periodic}.}
\label{enefreqspec}
\end{figure}

\hspace*{0.3cm}
The turbulent flow is composed of various vortical motions of different 
spatial and temporal scales. 
The dominant characteristic time-scale of turbulence is the 
large-eddy-turnover time $T_{\rm{\scriptscriptstyle T}}$, which 
may be estimated from the root-mean-square velocity and
the domain size, and is $\rm{O}(1)$ in the present flows. 
A more precise value of $T_{\rm{\scriptscriptstyle T}}$ 
may be obtained by the frequency spectra of energy 
$\mathcal{E}(t)$ and enstrophy $\mathcal{Q}(t)$, which will be 
useful for grouping of the periodic orbits studied in the next section. 
Time series of $\mathcal{E}(t)$ and $\mathcal{Q}(t)$, taken over 
$0<t <10^4$ in the turbulent flow at $\nu=0.0035$, 
are Fourier transformed, and their spectra are plotted in 
Fig. \ref{enefreqspec}. 
The dominant peak 
corresponds to the large-eddy-turnover time of 
$T_{\rm{\scriptscriptstyle T}}\approx 4.4$. 
The second peak near the left end shows variations on
time scales around $7T_{\rm{\scriptscriptstyle T}}$ and is not
discussed here. 
A weaker peak is visible at 
$T_{\rm{\scriptscriptstyle R}}\approx 2.2$, which 
corresponds to the period of oscillation of the flow observed at 
larger viscosity (see section \ref{sec:turbulent}) as well as to 
the most probable return time of the Poincar\'e map 
(see Fig. \ref{PDFt_r}) and will be used to label the periodic solutions. 

\section{Extracting periodic motion}
\label{sec:periodic}

\hspace*{0.3cm}
The state of the vorticity field is represented 
by a point in the
phase space spanned by $n$ Fourier components 
\{$\widetilde{\bm{\omega}}(\bm{k})$\} of the vorticiy field, independent 
under high-symmetry condition (\ref{hsymmetry}).
Here, $n$ is the number of degrees of freedom in the truncated system, 
about $10^4$ for $N=128$ as stated earlier. 
We specify an $(n-1)$-dimensional hyperplane $S$ 
by fixing one of 
the small wavenumber components of the vorticity field to a constant. 
Periodic orbits are then fixed points of $m$ iterations of 
Poincar\'e map $\mathcal{P}_{\nu}$ on $S$:
\begin{equation}
\mathcal{P}^{\ m}_{\nu}(\bm{y})-\bm{y}=\bm{0}\qquad \hbox{($m=1,2,3,\cdots$)},
\label{fixed}
\end{equation}
where $\bm{y}\in S$. 
Equation (\ref{fixed}) is highly
nonlinear and can be solved by Newton-Raphson iterations. For large $n$, 
the initial guess should be rather close to the fixed point to guarantee 
convergence.

\hspace*{0.3cm}
In order to find initial points, we performed a long time integration of 
Eqs. (\ref{NS}) and (\ref{cont}) with $\nu=0.0045$, i.e. 
in the weakly turbulent regime. We computed the intersection points with the 
plane $S$ given by $\widetilde{\omega}_1(0,2,4)=-0.04$, the time mean value
at $\nu=0.0045$.
If a point was mapped close to itself after $m$ iterations 
of the Poincar\'e map, i.e.
\begin{equation}
\| \mathcal{P}_{\nu}^{\ m}(\bm{y})-\bm{y} 
\|_{\rm{\scriptscriptstyle Q}} < \delta,
\end{equation}
it was marked as an initial point. 
Here, $\|\cdot\|_{\rm{\scriptscriptstyle Q}}$ 
stands for the enstrophy norm, i.e. 
the enstrophy computed according to Eq. (\ref{enstrdef}).
A suitable threshold value $\delta$ for the distance was given by $0.2$, 
about 10\% of the standard deviation of enstrophy.
Thus we found a collection of candidates for periodic orbits 
with $m$ ranging from $1$ to $12$. 
The same approach with $\nu =0.0035$, where turbulence was fully developed,
did not yield any candidates in a time integration of length 
$10^4 T_{\rm{\scriptscriptstyle T}}$.

\hspace*{0.3cm}
\begin{figure}[h]
\begin{center}
\includegraphics[width=.8\textwidth]{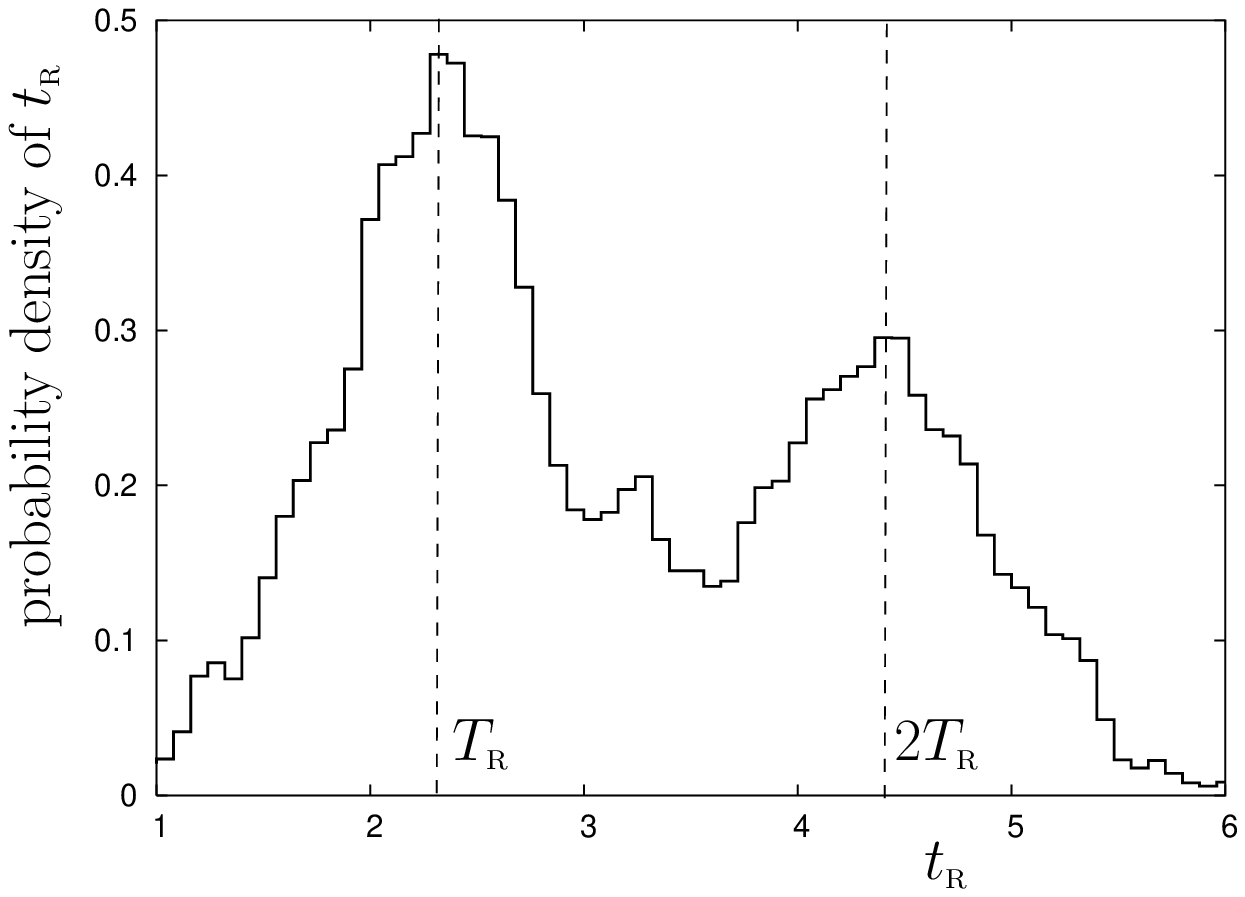}
\end{center}
\scaption{The probability density function 
of the return time $t_{\rm{\scriptscriptstyle R}}$ of the Poincar\'e map. 
Obtained from $2,000$ iterations of 
the Poincar\'e map at $\nu=0.0035$. The most probable
return time is $T_{\rm{\scriptscriptstyle R}}$. The probability density 
function shows little
dependence on the viscosity and the choice of the Poincar\'e plane $S$. 
}
\label{PDFt_r}
\end{figure}

\hspace*{0.3cm}
Fig. \ref{PDFt_r} shows the probability density function 
of the return time $t_{\rm{\scriptscriptstyle R}}$ of the Poincar\'e
map, computed at $\nu=0.0035$. 
Two large peaks are prominent around $T_{\rm{\scriptscriptstyle R}}$ and 
$2T_{\rm{\scriptscriptstyle R}}$. 
This implies that $\widetilde{\omega}_1(0,2,4)$ oscillates with frequency 
about $T_{\rm{\scriptscriptstyle R}}$ and that it crosses the prescribed value 
$-0.04$ every oscillation with occasional missing of a crossing. 
Two and more successive missings are very rare. 
Recall that the most probable return time $T_{\rm{\scriptscriptstyle R}}$
is the same as the characteristic time of turbulence 
identified in section \ref{sec:turbulent} as a peak in the 
frequency spectra of energy and enstrophy. 
The probability density function shows little dependence on the viscosity.
The periodic orbits identified as
fixed points of $\mathcal{P}^{\ m}_{\nu}$ have a period roughly equal to 
$m$ times $T_{\rm{\scriptscriptstyle R}}$ in the whole 
range $0.0035<\nu<0.0045$.
In the following we refer to them as period-$m$ orbits and denote 
their period by $T_{m{\rm p}}$.

\hspace*{0.3cm}
From the periodic orbits found in the weakly turbulent regime, we select 
orbits with periods $1$ up to $5$ and continue them down to
$\nu=0.0035$. 
For continuation of the periodic orbits, we use the arc-length 
method, a prediction-correction method which requires solving
an equation similar to Eq. (\ref{fixed}) at each continuation step. 
The most time-consuming part of this algorithm is
the computation of derivatives  
$\mbox{D}_{{\bm{\scriptstyle y}},\nu}\mathcal{P}_{\nu}$ 
of the Poincar\'e map 
with respect to the $(n-1)$ components of $\bm{y}$ and $\nu$. 
Finite differencing is employed for the derivatives so that 
for each Newton-Raphson iteration we have to run $(n+1)$ integrations, 
which can conveniently be done in parallel. We use $128$ processors 
simultaneously on a Fujitsu GP7000F900 parallel computer. 
The computation of one iteration of the Poincar\'e
map and its derivatives takes about $25$ minutes of CPU time on each 
processor. 
The average step size in the parameter is $\Delta \nu \approx 0.00004$ and 
about three Newton-Raphson iterations are taken at each continuation step 
before the residue is smaller than $10^{-9}$ in the
enstrophy norm. This brings the total computation time for continuation
of a period one ($m=1$) orbit down to $\nu=0.0035$ to about $31$ hours.
Note that there is no guarantee that an orbit can be
continued all the way. In fact, about half the continuations we ran ended
in a bifurcation point before reaching the maximal micro-scale Reynolds 
number. 

\begin{figure}[t]
\begin{center}
\includegraphics[width=1.0\textwidth]{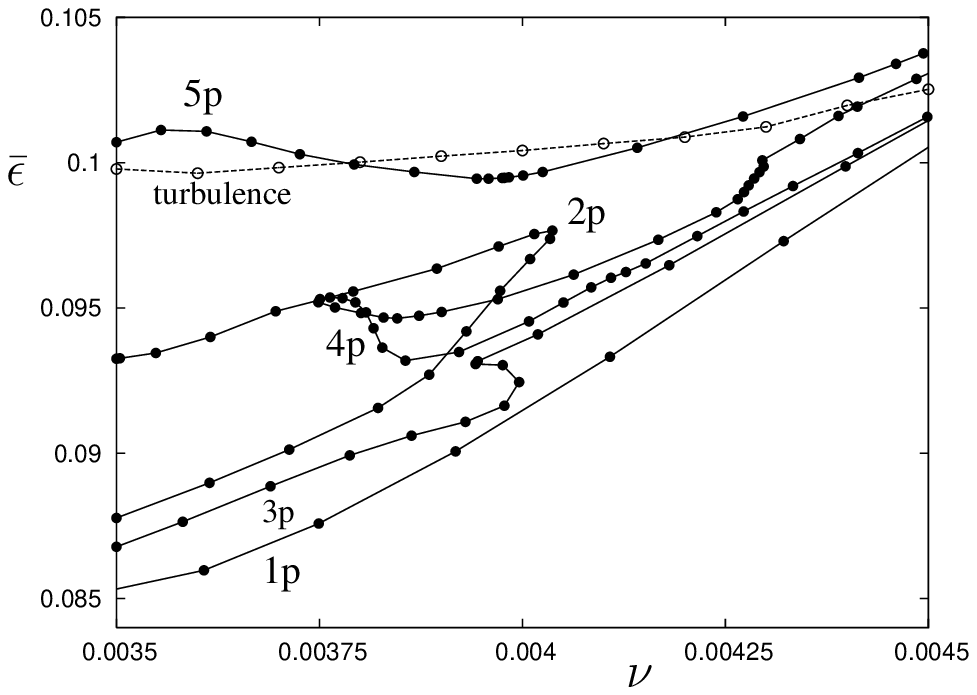}
\end{center}
\scaption{Energy-dissipation rate averaged over 
the periodic orbits as a function of viscosity. 
The label $m$p of the individual curves indicates an orbit corresponding to a 
fixed point of $\mathcal{P}^{\ m}_{\nu}$ 
and having a period roughly equal to $mT_{\rm{\scriptscriptstyle R}}$. 
The dotted line denotes the values in the turbulent state.}
\label{continuation}
\end{figure}

\hspace*{0.3cm}
It is our primary concern to find out whether the periodic orbits
may represent the turbulent state or not.
For this purpose we compute the mean energy-dissipation rate 
$\overline{\epsilon}$, averaged along the periodic orbits
at each point on the continuation curve, 
and compare these values to that of the 
turbulent state. 
As seen in the preceding section, 
the time-averaged energy-dissipation 
rate $\overline{\epsilon}$ tends to saturate around 
$0.1$ in the turbulent state for $\nu < 0.004$ 
(see Fig. \ref{turbulent1}). 
In Fig. \ref{continuation}, we compare $\overline{\epsilon}$ averaged over 
the periodic motion to that for the turbulent state.
Clearly, 
the values given by the short periodic orbits diverge from that 
of the turbulent state, decreasing monotonically with viscocity. 
The value produced by the period-5 orbit,
however, stays close to the one found for the turbulent state. 

\hspace*{0.3cm}
The period-2 orbit is the only one found at a somewhat lower viscosity, 
$\nu=0.004$, by the method described above.
At the time of writing, the continuation curves for the period-3 and 
period-4 solutions were incomplete. 
These continuations are currently running on a shared memory system
which is considerabaly slower than the 128 CPU parallel machine. 

\section{Embedded periodic motion}
\label{sec:embedded}

\hspace*{0.3cm}
The results of the preceeding section suggest that the period-5 orbit  
represents the turbulent state. 
We now analyse the properties of this orbit 
for $\nu=0.0035$ in detail. 

\subsection{Structure in Phase Space}
\label{subsec:structure}

\begin{figure}[h]
\begin{center}
\includegraphics[width=1.0\textwidth]{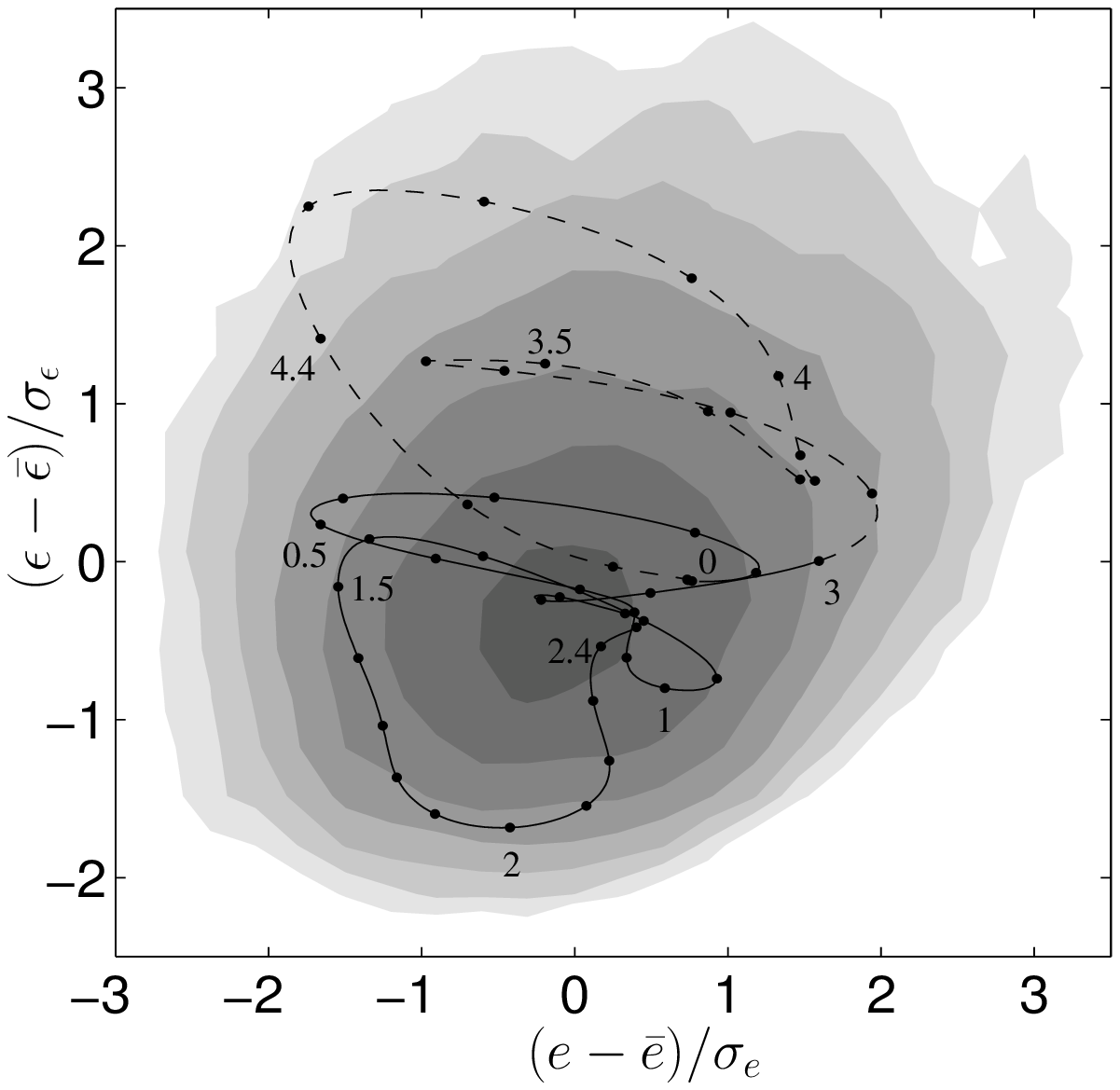}
\end{center}
\scaption{The period-5 orbit and the probability density function of 
the turbulent state projected on the ($e$, $\epsilon$)-plane 
at $\nu=0.0035$. 
The periodic orbit of period 
$T_{5{\rm p}}=4.91T_{\rm{\scriptscriptstyle R}}$ 
is represented by a closed curve with 
solid ($t/T_{\rm{\scriptscriptstyle R}}<3$) and 
dashed ($t/T_{\rm{\scriptscriptstyle R}}>3$)
parts in the low-activity and high-activity periods, respectively. 
Dots are attached at every $0.1 T_{\rm{\scriptscriptstyle R}}$.  
Numbers indicates the time in the unit of $T_{\rm{\scriptscriptstyle R}}$. 
Contours of the probability density function are drawn at $80\%$ of its 
peak value and successive factors of $0.5$ with larger values in darker 
areas.
On the axes the energy input rate and dissipation rate are shown as 
deviations from 
their temporal mean normalised by their standard deviation in turbulence.}
\label{eineout}
\end{figure}

\hspace*{0.3cm}
It is impossible to show 
how close this orbit is to the turbulent state 
in the $n$-dimensional phase space, but we can get an impression
by looking at
its projection on the two-dimensional $(e, \epsilon)$-plane 
spanned by the energy-input rate and the energy-dissipation rate. 
In Fig. \ref{eineout}, we plot this projection of the orbit 
for $\nu=0.0035$ by a closed curve 
with dots at every $0.1T_{\rm{\scriptscriptstyle R}}$. 
The solid and dashed parts of the curve respectively indicate the low-activity 
and high-activity periods described below. 
Numbers attached to the orbit are measured from an arbitrary reference 
time near the beginning of the low-activity period and normalised 
by the return time $T_{\rm{\scriptscriptstyle R}}$. 
Contours of grey scale show the probability density function of the 
turbulent state with larger values in darker areas. 
Both axes are normalised by the standard deviation around the 
temporal mean of the respective quantitites in turbulence. 

\begin{figure}[tbp]
\begin{center}
(a)
\includegraphics[width=0.9\textwidth]{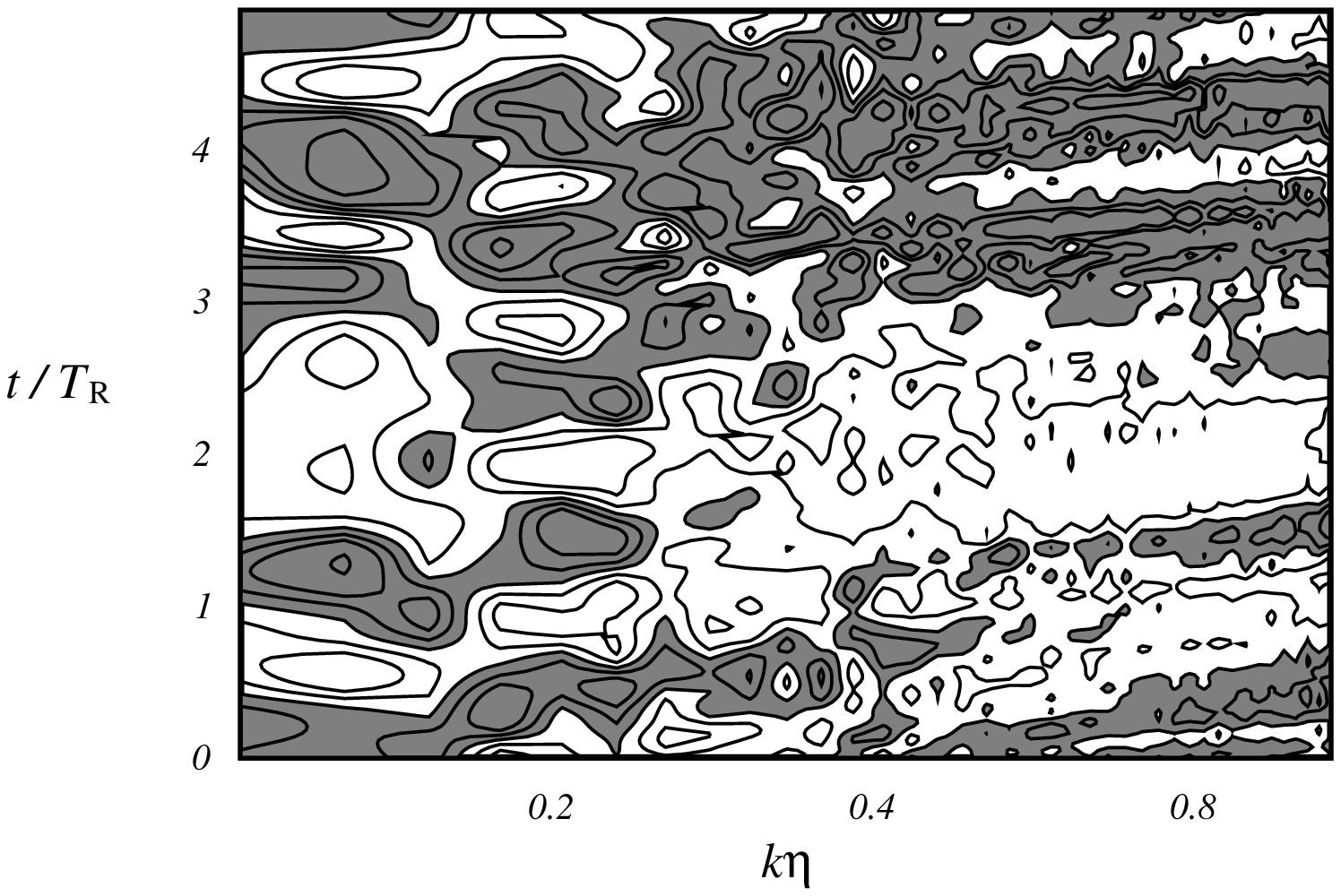}

\vspace{0.5cm}

(b)
\includegraphics[width=0.9\textwidth]{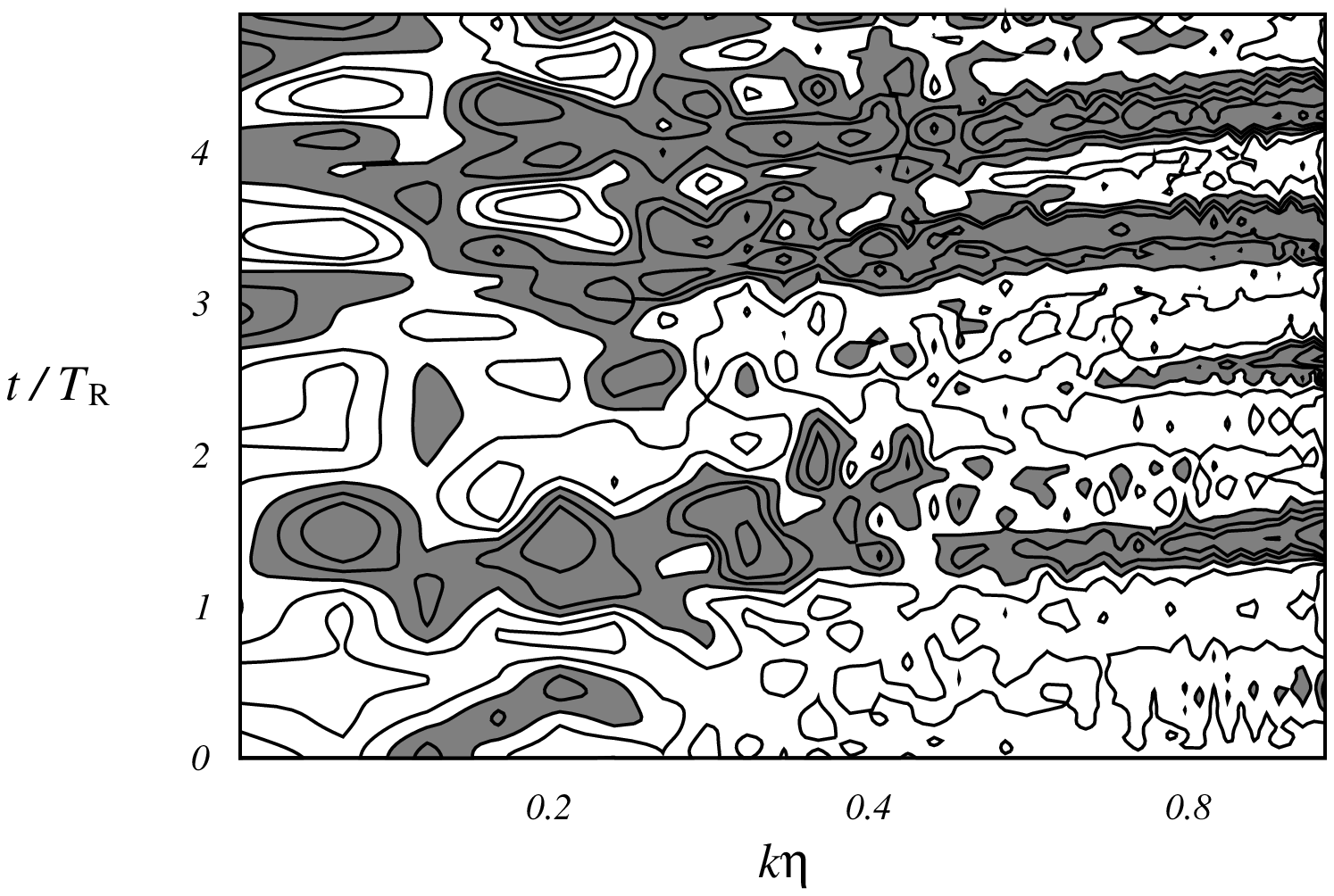}
\end{center}
\scaption{Temporal evolution of energy spectrum. 
The excess from temporal mean normalised by standard deviation 
of the three-dimensional energy spectrum $E(k,t)$ is shown by 
contours at the levels of $0$, $\pm 0.25$, $\pm 0.5$, $\pm 1$, $\pm 2$, 
the positive parts being shaded.  
The abscissa is logarithm of the wavenumber normalised by 
Kolmogorov length $\eta$ and
the ordinate is the time normalised by $T_{\rm{\scriptscriptstyle R}}$. 
(a) period-5 orbit. (b) Fragment from 
a turbulent time series. 
The tilted streaks represent anomalies cascading into the dissipation 
range.}
\label{kshells}
\end{figure}

\hspace*{0.3cm}
This figure has several interesting features.
First, the probability density function of turbulent state is slightly 
skewed towards high $e$ and high $\epsilon$, and the peak 
is located at the lower-left side of the origin. 
This is due to bursting events in which anomalous 
amounts of kinetic energy are injected and dissipated. 
The periodic orbit makes a large excursion to high-$\epsilon$ 
corresponding to such a burst 
Secondly, the distance of the periodic orbit from the origin remains of the
order of the standard deviations of turbulence.
This is consistent with the picture that this orbit is embedded in the 
turbulent state. 
In fact, both the mean values and the standard devations of $e(t)$ 
and $\epsilon(t)$ are strikingly close for the turbulence and the
periodic motion; namely, they are 
$\overline{e}=\overline{\epsilon}=0.0998$, 
$\sigma_{e}=0.0352$, $\sigma_{\epsilon}=0.0155$ for the former and 
$\overline{e}^{5{\rm p}}=\overline{\epsilon}^{5{\rm p}}=0.107$, 
$\sigma_{e}^{5{\rm p}}=0.0348$, 
$\sigma_{\epsilon}^{5{\rm p}}=0.0141$ for the latter. 
The standard deviation of $e(t)$ is larger than that of $\epsilon(t)$ 
by about factor 2. 
The magnitude of fluctuations of the present turbulence 
is about 35\% of the mean values in the energy-input rate 
($\sigma_{e}/\overline{e}=0.353$),
and 16\% in the energy-dissipation rate 
($\sigma_{\epsilon}/\overline{\epsilon}=0.156$). 
Thirdly, although the trajectory of the periodic orbit is not simple, 
we can see that it generally rotates counter-clockwise. 
In other words, 
peaks of $\epsilon(t)$ come after those of $e(t)$, which is consistent with 
the picture of energy cascade to larger wavenumbers (see Fig. \ref{kshells}). 
Fourthly, the orbit may be divided into two periods. 
During the first period (solid line) 
of about $3T_{\rm{\scriptscriptstyle R}}$, $e(t)$ and
$\epsilon(t)$ are near or below their mean values. 
This is the period of low activity. 
It is followed by a period of high activity 
(dashed line) of about 
$2T_{\rm{\scriptscriptstyle R}}$. 
Thus, the transitions between the low
activity and the high activity phase take place on a time scale 
$T_{\rm{\scriptscriptstyle T}}$, the large-eddy-turnover time.
Dots are drawn on the periodic orbit at equal time intevals so that we can 
get an impression of the speed of the state point along the orbit, which
tends to be higher during the high activity phase. 

\begin{figure}[t]
\begin{center}
(a)
\includegraphics[width=0.7\textwidth]{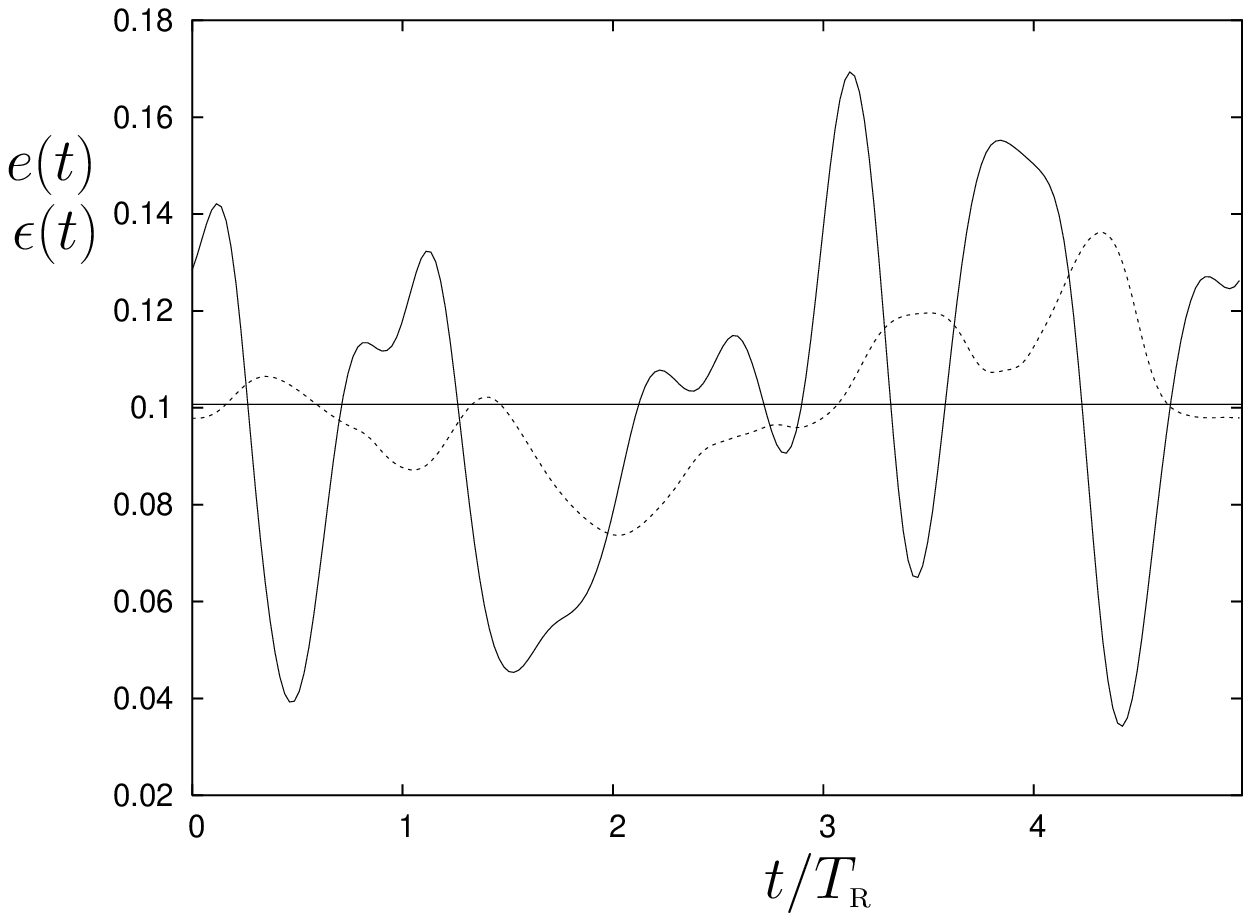}

\vspace*{1.0cm}

(b)
\includegraphics[width=0.7\textwidth]{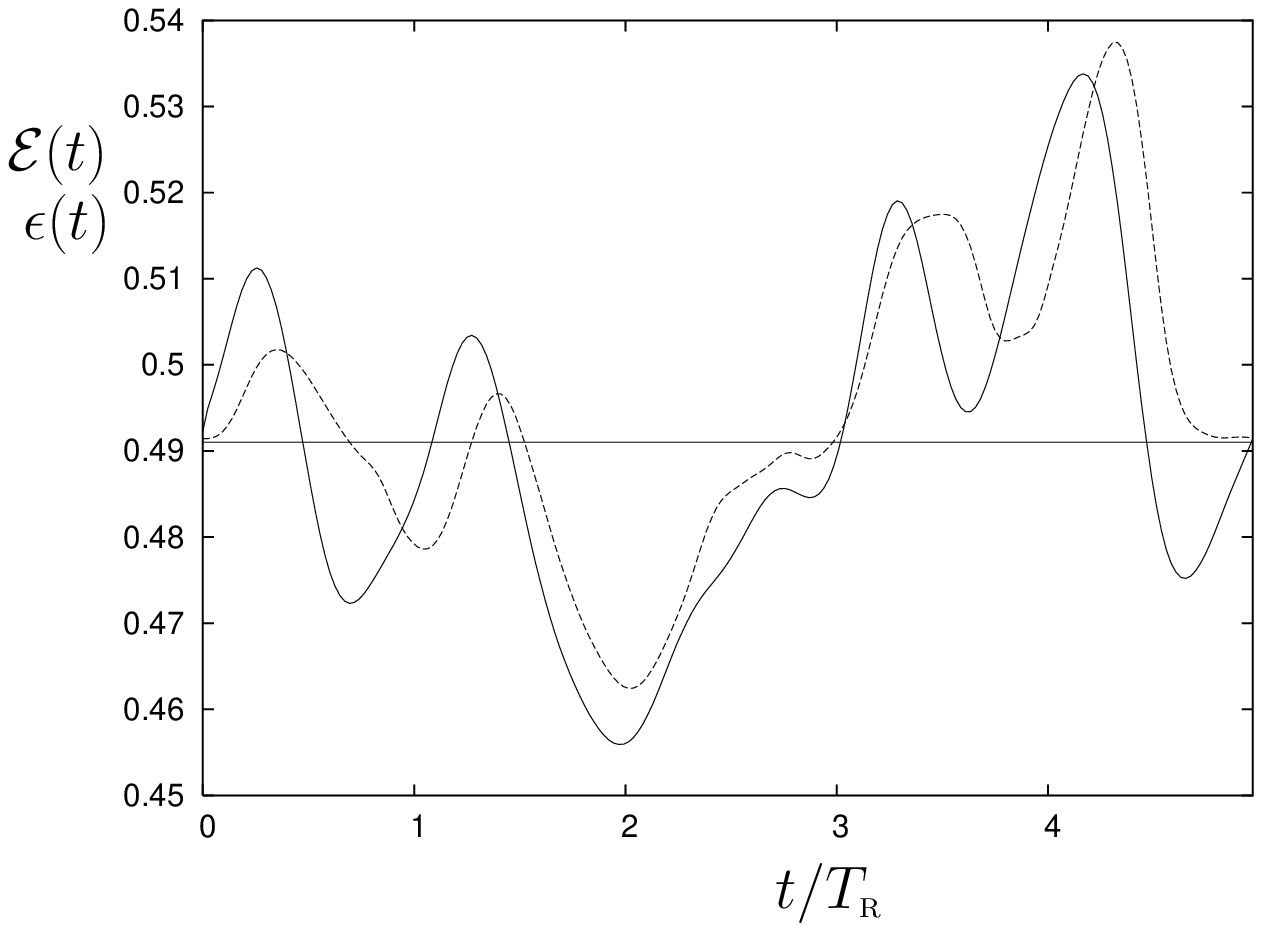}
\end{center}
\scaption{Energy characteristics of the period-5 orbit. 
Temporal variations of (a) energy-input rate $e(t)$ (solid line) and 
energy-dissipation rate $\epsilon(t)$ (dashed line) and (b) energy $\mathcal{E}(t)$ 
(solid line) and $\epsilon(t)$ (dashed line), the latter normalised to have the same time mean. 
The horizontal lines indicate the mean values of the respective 
quantities. 
The abscissa is the time normalised by $T_{\rm{\scriptscriptstyle R}}$. 
}
\label{fig:time-series}
\end{figure}

\hspace*{0.3cm}
The time series of $e(t)$ and $\epsilon(t)$, shown in Fig.
\ref{fig:time-series}(a) is another representation of the periodic orbit. 
We see that this periodic motion is composed of five 
enhanced actions of energy input and dissipation every 
$T_{\rm{\scriptscriptstyle R}}$. 
The input rate is stronger in amplitude than the dissipation rate.
The oscillation phase is anti-correlated between the two. 
It is clearly seen from the behaviour of $\epsilon(t)$ that 
the periods of low activity and high activity are the intervals of 
$t/T_{\rm{\scriptscriptstyle R}}<3$ and 
$t/T_{\rm{\scriptscriptstyle R}}>3$, respectively. 
Both $\epsilon(t)$ and $e(t)$ oscillate around 
lower (or higher) values than their mean values 
(denoted by the horizontal line) 
in the former (or latter) interval. 
In Fig. \ref{fig:time-series}(b) is shown the time series of energy 
$\mathcal{E}(t)$, the time-derivative of which is equal to the difference 
$e(t)-\epsilon(t)$. 
For comparison, the time series of $\epsilon(t)$ is also plotted 
after shifting and scaling appropriately. 
It is interesting that the energy and energy-dissipation rate change 
quite similarly and that the peaks of the former proceed a 
little those of the latter. 
Furthermore, comparison with Fig. \ref{fig:time-series}(a) tells us 
that peaks of $e(t)$ precede those of $\mathcal{E}(t)$. 
This order of the peaks represents the energy cascade process. 

\begin{figure}[h]
\begin{center}
\includegraphics[width=0.9\textwidth]{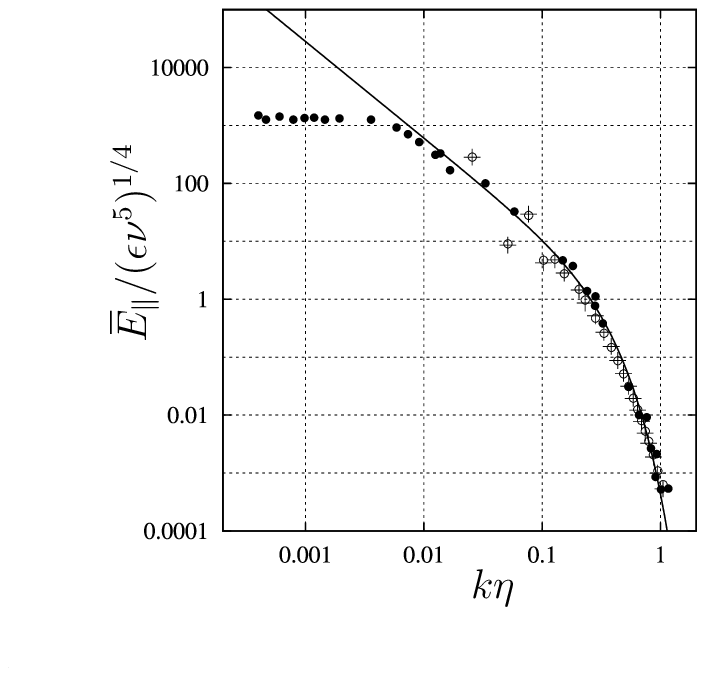}
\end{center}
\scaption{One-dimensional longitudinal energy spectrum. 
Open circles and pluses represent respectively the turbulent state and 
the period-5 motion in high-symmetric flow at $R_{\lambda}=67$. 
Solid circles denote experimental data at $R_{\lambda}=130$ 
taken from \citet{champ}. 
The solid line represents 
the asymptotic form at 
$R_{\lambda}\rightarrow \infty$ derived theoretical by the sparse 
direct-interaction approximation \cite{KiGo97}.
The axes are normalised according to the Kolmogorov scaling.}
\label{1Dspectra2}
\end{figure}

\hspace*{0.3cm}
Another convenient projection to capture the structure 
of the periodic orbit in the phase space is given by taking an 
arithmetic average of the square of those components of 
\{$\widetilde{\bm{\omega}}(\bm{k})$\} that have the same magnitude of 
wavenumber $|\bm{k}|$. 
This is nothing but the enstrophy spectrum, identical to the 
energy spectrum multiplied by the wavenumber squared.  
Among others, 
the one-dimensional longitudinal energy spectrum can readily be compared to 
laboratory experiments. 
In Fig. \ref{1Dspectra2}, we plot the time-averaged energy spectrum 
of our simulations 
with open circles for turbulence and with pluses for the period-5 
motion. 
For comparison, also shown are the laboratory data in shear flow at 
$R_{\lambda}=130$ with solid circles \cite{champ} and the asymptotic form at 
the infinite Reynolds number derived theoretically with a solid line 
\cite{KiGo97}.
It is remarkable that the data of the periodic motion and turbulence 
agree with each other almost perfectly, providing us with another 
support of closeness in the phase space of this periodic motion and 
turbulence. 
The data nicely collapse onto a single curve beyond the energy containing 
range, which shows that the agreement between the spectra of periodic and
turbulent motion is not an artifact of the high symmetry of the 
present numerical flow. 

\hspace*{0.3cm} 
So far we have seen that the period-5 motion reproduces the 
temporal mean energy spectrum of turbulent state remarkably well. 
A yet more detailed comparison between the period-5 
motion and turbulence is provided by the temporal evolution of the energy 
spectral function. 
In Fig. \ref{kshells}(a), we show the time series of the three-dimensional 
energy spectral function calculated as
\begin{equation}
E(k,t)=\frac{1}{2}
\sum_{k-\frac{1}{2}\le |\bm{\scriptstyle k}'|<k+\frac{1}{2}}
|\widetilde{\bm{u}}(\bm{k}',t)|^2.
\label{eq:3D}
\end{equation}
In order to emphasize the fluctuations, the departure from the 
temporal mean, 
normalised by the standard deviation of the spectrum, is plotted by contours 
with positive parts shaded.
The abscissa, the wavenumber normalised by the Kolmogorov length, 
is scaled logarithmically to illuminate the cascade process, 
thought to be a series of breakdowns of coherent vortical structures 
into parts about half their size. 

\hspace*{0.3cm}
It is not straightforward to compare the pattern of $E(k,t)$ of the 
periodic orbit shown in 
Fig. \ref{kshells}(a) to that of turbulence 
because we do not know {\it a priori} 
which parts of a turbulent time sequence are to be compared with.
We can, however, select  
portions of a time series of turbulence that are close to 
the periodic orbit, as will be explained in the next subsection. 
In Fig. \ref{kshells}(b), we show such a portion of a time series of 
$E(k,t)$ of turbulence over the same 
time interval as the periodic motion. The time variable is the same as in 
Figs. \ref{eineout} and \ref{fig:time-series}. 
The pattern of the energy spectrum of Figs. \ref{kshells}(a) and (b) 
is remarkably similar, which adds to the evidence that this period-5 
orbit represents the turbulent state well. 
Note especially the
inclination and mutual spacing of the streaks which show the cascade
process and the relative duration of the 

\noindent
low-activity period ($t/T_{\rm{\scriptscriptstyle R}}<3$) and 
the high-activity period ($t/T_{\rm{\scriptscriptstyle R}}>3$).
Two wide streaks at larger wavenumbers ($k\eta > 0.4$) 
in the later phase of the periodic motion correspond to the 
excursion to high $\epsilon$ seen in Fig. \ref{eineout}, whereas 
the other narrow streaks in the earlier phase to the slow excursion. 
Initial conditions corresponding to other local minima give similar 
pictures. 
See \citet{kida3} for a detailed discussion on the energy dynamics at 
a larger Reynolds number $R_{\lambda}=186$. 

\subsection{Periodic motion as the skeleton of Turbulence}

\hspace*{0.3cm} 
Motivated by the speculation that the turbulent state approaches
the periodic orbit frequently, we introduce
a measure of closeness by the 
`distance' $D(t)$, in the 
$(e,\epsilon)$ plane, 
between the periodic orbit and a finite, turbulent time sequence of lenght 
$T_{\rm{\scriptscriptstyle T}}$ as
\begin{eqnarray}
D(t)^2=\frac{1}{2T_{\rm{\scriptscriptstyle T}}}
\min_{0\leq t^* <T_{5\rm{\scriptscriptstyle p}}}
\int_{-\frac{1}{2}T_{\rm{\scriptscriptstyle T}}}
^{\frac{1}{2}T_{\rm{\scriptscriptstyle T}}}&&
\Big[\frac{1}{\sigma_{\epsilon}^2}
(\epsilon^{5{\rm p}}(t^*+t')-\epsilon(t+t'))^2\nonumber\\[0.2cm]
&&+\frac{1}{\sigma_{e}^2}(e^{5{\rm p}}(t^*+t')
-e(t+t'))^2\Big]\mbox{d}t',
\label{D}
\end{eqnarray}

\vspace*{0.2cm}

\begin{figure}[h]
\begin{center}
\includegraphics[width=0.8\textwidth]{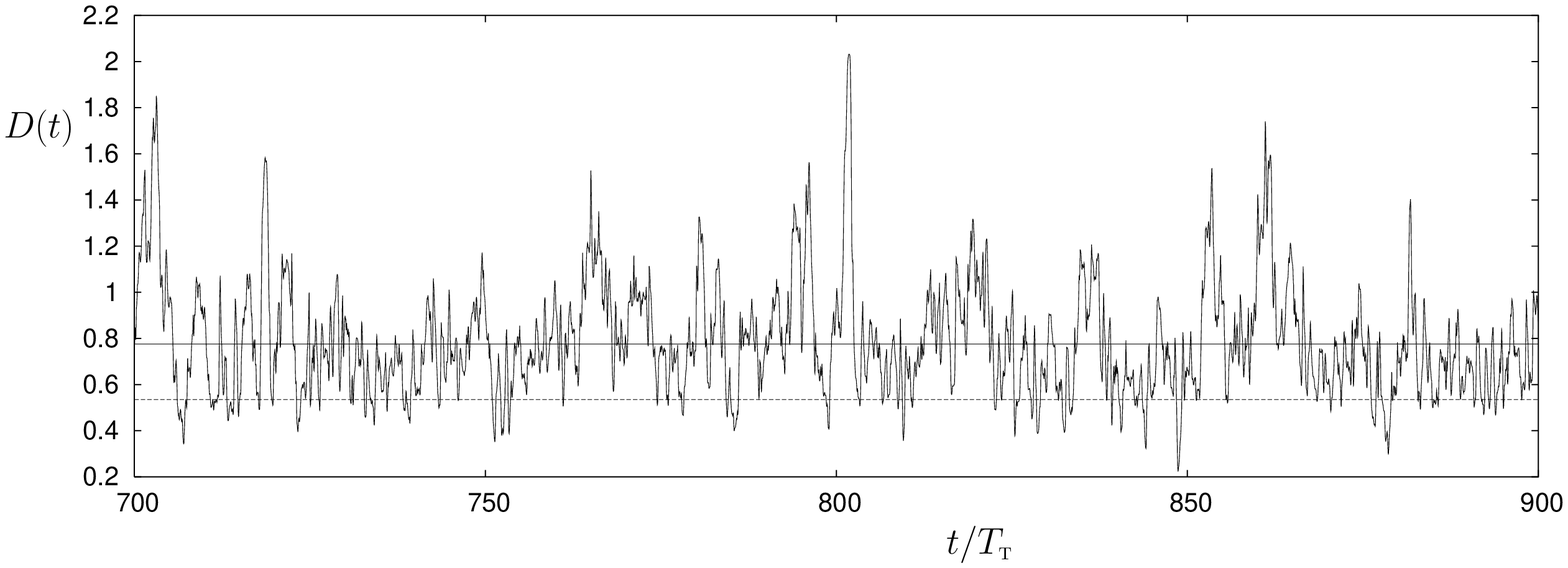}
\end{center}
\scaption{Temporal variation of the distance $D(t)$ between the turbulent 
state and the period-5 motion over an arbitrary time interval. 
The horizontal lines denote the time mean $\overline{D}$ (solid)
and $\overline{D}-\sigma_{\rm{\scriptscriptstyle D}}$ (dashed).
The abscissa is the time normalised by large-eddy-turnover time 
$T_{\rm{\scriptscriptstyle T}}$. 
Observe that the frequency of appearance of sharp minima is of 
$O(T_{\rm{\scriptscriptstyle T}})$. 
}
\label{fig:distance}
\end{figure}

\begin{figure}[t]
\begin{center}
\includegraphics[width=0.75\textwidth]{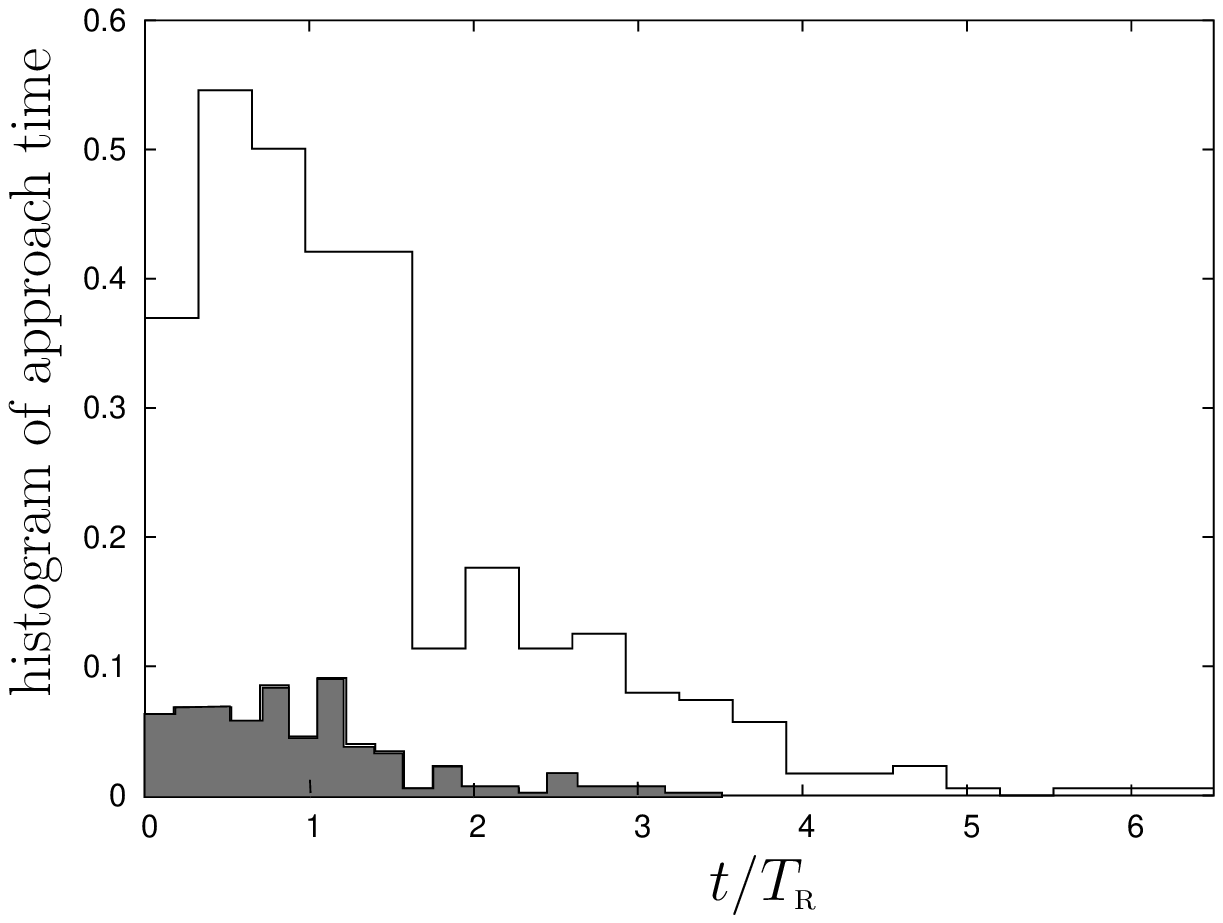}
\end{center}
\scaption{Frequency distribution of approach time of turbulent state 
to the period-5 motion. 
The histgrams of time during which 
$D(t) < \overline{D}-\sigma_{\rm{\scriptscriptstyle D}}$ 
and $D(t) < \overline{D}-1.5\sigma_{\rm{\scriptscriptstyle D}}$ 
are shown by white and grey steps, respectively. 
The area of the former histgram is normalised to unity. 
The mean values of the approach time are $1.33 T_{\rm{\scriptscriptstyle R}}$
($=0.66T_{\rm{\scriptscriptstyle T}}$) and 
$0.91 T_{\rm{\scriptscriptstyle R}}$ ($=0.45 T_{\rm{\scriptscriptstyle T}}$)
for the respective cases. 
}
\label{fig:approach-time}
\end{figure}

\noindent
where $\epsilon^{5{\rm p}}(t)$ and 
$e^{5{\rm p}}(t)$ denote the energy-dissipation rate and energy-input 
rate along the orbit, respectively. 
This distance is normalised such that, if wereplace 
$\epsilon^{5{\rm p}}(t)$ and 
$e^{5{\rm p}}(t)$ by their respective mean values in the turbulent state, 
the temporal mean of $D(t)$ is unity. 
A time series of $D(t)$ taken from a long integration is shown in 
Fig. \ref{fig:distance} together with the temporal mean 
$\overline{D}$ ($=0.776$) 
and the temporal mean minus the standard deviation 
$\overline{D}-\sigma_{\rm{\scriptscriptstyle D}}$ ($=0.535$). 
It can be seen that $D(t)$ takes sharp minima at the rate of once 
over the period of $O(T_{\rm{\scriptscriptstyle T}})$, implying 
that the turbulent state approaches the 
period-5 orbit at intervals of about one large-eddy-turnover time. 

\hspace*{0.3cm} 
In order to discuss the approach frequency of the turbulent state 
to the period-5 orbit quantitatively, we consider the statistics of 
intersections between $D(t)$ and the two horizontal lines. 
We may say that the turbulent state is located within $\overline{D}$ 
(or $\overline{D}-\sigma_{\rm{\scriptscriptstyle D}}$) distance 
from the periodic orbit when $D(t) < \overline{D}$ 
(or $D(t) < \overline{D}-\sigma_{\rm{\scriptscriptstyle D}}$). 
The intervals between two consecutive intersection 
times with $D(t)$ below the holizontal lines are called the 
approach time, which are regarded as 
the periods when the turbulent state are close to the periodic orbit. 
The approach time is different depending on the threshold distance. 
In Fig. \ref{fig:approach-time}, we plot their histgrams, 
obtained from a time series of about $7,000$ non-normalised 
time units, for two threshold distances, 
$\overline{D}-\sigma_{\rm{\scriptscriptstyle D}}$ (white steps) and 
$\overline{D}-1.5\sigma_{\rm{\scriptscriptstyle D}}$ (grey steps). 
The area is normalised to be unity for the former histogram. 
The mean approach times are 
$1.33 T_{\rm{\scriptscriptstyle R}}$ 
($0.66 T_{\rm{\scriptscriptstyle T}}$) and 
$0.91 T_{\rm{\scriptscriptstyle R}}$ 
($0.45 T_{\rm{\scriptscriptstyle T}}$) for 
the respective thresholds, implying that the turbulent state is likely to 
stay around the period-5 orbit over the time of 
$\text{O}(T_{\rm{\scriptscriptstyle T}})$ every time it approaches. 

\begin{figure}[t]
\begin{center}
\includegraphics[width=0.75\textwidth]{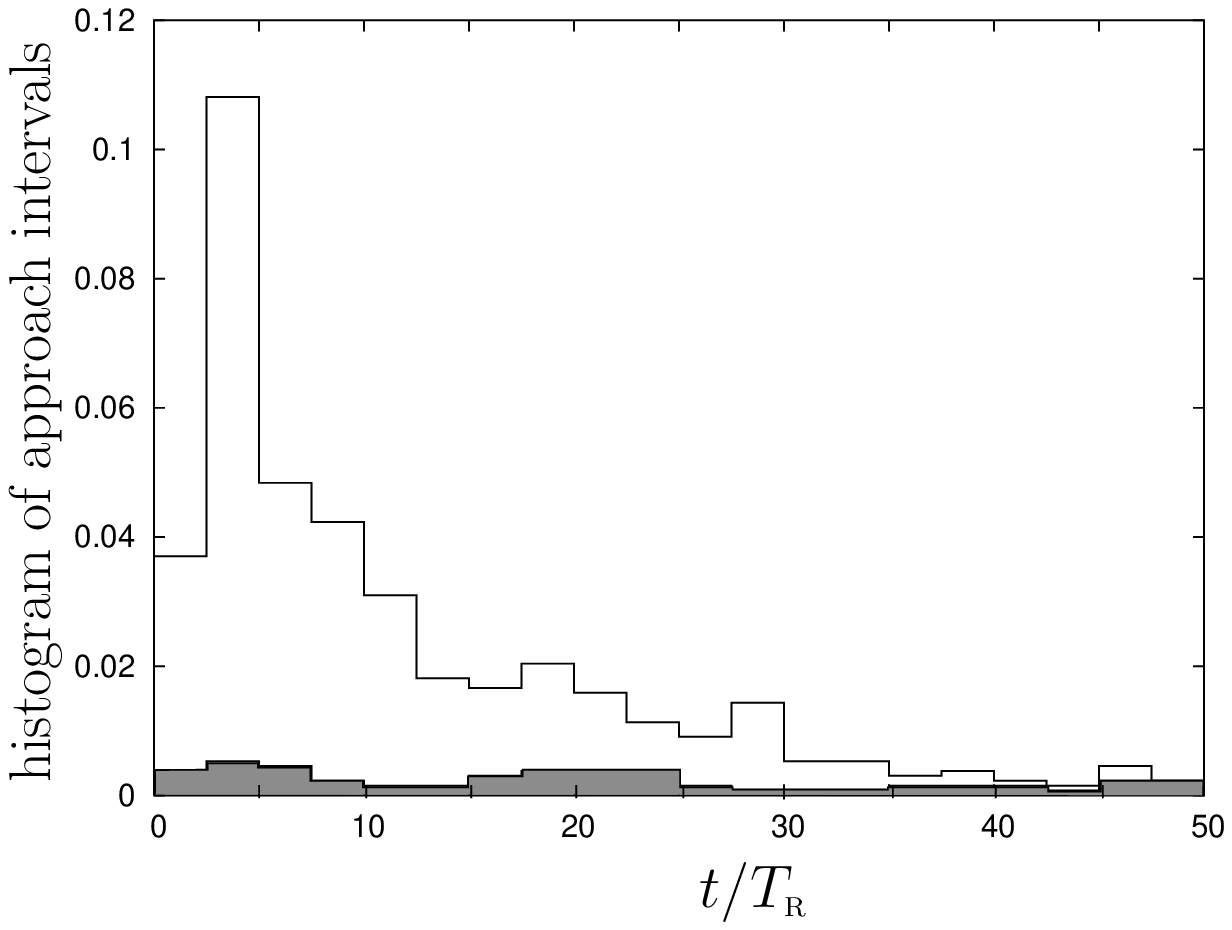}
\end{center}
\scaption{Frequency distribution of approach intervals of turbulent 
state to the period-5 motion. 
The histgrams of time interval between midpoints of segments with 
$\overline{D}-\sigma_{\rm{\scriptscriptstyle D}}$ and 
$\overline{D}-1.5\sigma_{\rm{\scriptscriptstyle D}}$ are shown by 
white and grey steps, respectively. 
The area of the former histgram is normalised to unity. 
The mean values of the interval is $5.8T_{\rm{\scriptscriptstyle R}}$ 
($=2.9T_{\rm{\scriptscriptstyle T}}$) and 
$28.0 T_{\rm{\scriptscriptstyle R}}$ ($=14.0 T_{\rm{\scriptscriptstyle T}}$) for the respective segments. 
}
\label{fig:approach-interval}
\end{figure}

\hspace*{0.3cm} 
How frequently the turbulent state approaches the periodic orbit 
may be measured by the time intervals between consecutive approach 
periods, which is called the approach interval. 
Note that this measure is more appropriate than counting the local 
minimum times of $D(t)$ in Fig. \ref{fig:distance} because two or 
more minima may occur in one approach interval. 
In Fig. \ref{fig:approach-interval}, we plot the histgrams of the 
approach interval made by using the same data as that for 
Fig. \ref{fig:approach-time}. 
Again, the white and grey steps indicate the histgrams for the threshold 
distances of $\overline{D}-\sigma_{\rm{\scriptscriptstyle D}}$ and 
$\overline{D}-1.5\sigma_{\rm{\scriptscriptstyle D}}$, respectively. 
The area is normalised to be unity for the former one. 
The mean approach intervals are 
$5.8 T_{\rm{\scriptscriptstyle R}}$ ($=2.9 T_{\rm{\scriptscriptstyle T}}$) 
and $28.0 T_{\rm{\scriptscriptstyle R}}$ 
($=14.0 T_{\rm{\scriptscriptstyle T}}$)
for the respective thresholds. 
This result tells us that the turbulent state approaches the 
period-5 orbit at the rate of once over a few eddy-turnover times 
(or about the period of this periodic orbit) within distance of 
$\overline{D}-\sigma_{\rm{\scriptscriptstyle D}}$ and that 
a more closer approach within distance of 
$\overline{D}-1.5\sigma_{\rm{\scriptscriptstyle D}}$ is 
observed much less frequently, i.e. once over $14$
eddy-turnover times. 

\begin{figure}[t]
\begin{center}
\includegraphics[width=0.75\textwidth]{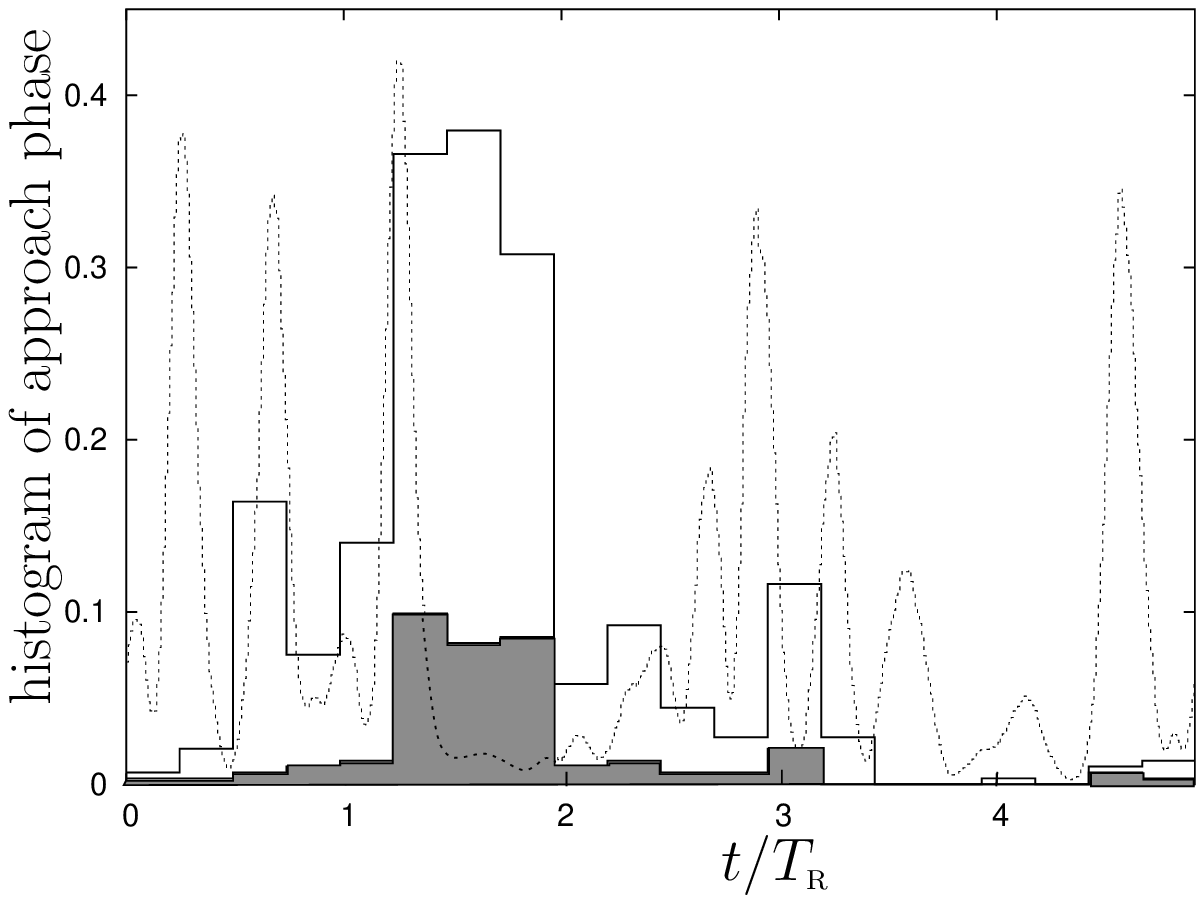}
\end{center}
\scaption{Frequency distribution of the phase of the period-5 motion 
at which the turbulent state approaches. 
The histgrams of the phase at the mid-point of segments with  
$D(t) < \overline{D}-\sigma_{\rm{\scriptscriptstyle D}}$ 
and $D(t) < \overline{D}-1.5\sigma_{\rm{\scriptscriptstyle D}}$ 
are shown by white and grey steps, respectively. 
The area of the former histgram is normalised to unity. 
The dotted curve indicates the existing probability 
of the turbulent state on the period-5 orbit projected on 
the $(e, \epsilon)$-plane. 
}
\label{fig:phase}
\end{figure}

\hspace*{0.3cm} 
Which parts of the periodic orbit is the turbulent state 
likely to approach more frequently ? 
This information is provided by the phase time $t^*$ that 
defines the distance $D(t)$, i.e. that gives the minimum value of 
the integration in (\ref{D}). 
In Fig. \ref{fig:phase}, we show the histgrams of the phase time of 
approach of turbulent state for threshold distances of 
$\overline{D}-\sigma_{\rm{\scriptscriptstyle D}}$ (white steps) and 
$\overline{D}-1.5\sigma_{\rm{\scriptscriptstyle D}}$ (grey steps). 
The area is normalised to be unity for the former histgram. 
For comparison, the PDF of realisation of the turbulent state 
along the period-5 orbit is drawn with a dotted curve.
This density is obtained by integrating the PDF shown in Fig. \ref{eineout}
over a small neigbourhood (a disk with a radius much smaller than the 
standard deviations $\sigma_{e}$ and $\sigma_{\epsilon}$)  of a given point
on the period-5 orbit and multiplying by the local speed of the state point. 
It is interesting that the approach phase is localised 
in the low-active period ($t^*/T_{\rm{\scriptscriptstyle R}}< 3$), 
but hardly observed in 
the high-active period ($t^*/T_{\rm{\scriptscriptstyle R}}> 3$). 
This tendency of non-uniform appoach suggests that the movement of 
the state point of turbulence may be more violent in the high-active 
period than in the low-active period. 
The stability characteristics, in the phase space, of the state point 
will be examined by the local Lyapunov analysis in the next subsection. 
Incidentally, the totally different behaviour between the histgrams 
(steps) of the approach phase and the existing probability (dotted curve) 
of turbulent state indicates that the non-uniformity of the approach 
phase may be due to that of the dynamical properties 
along the periodic orbit. 

\begin{figure}[h]
\begin{center}
\includegraphics[width=0.48\textwidth]{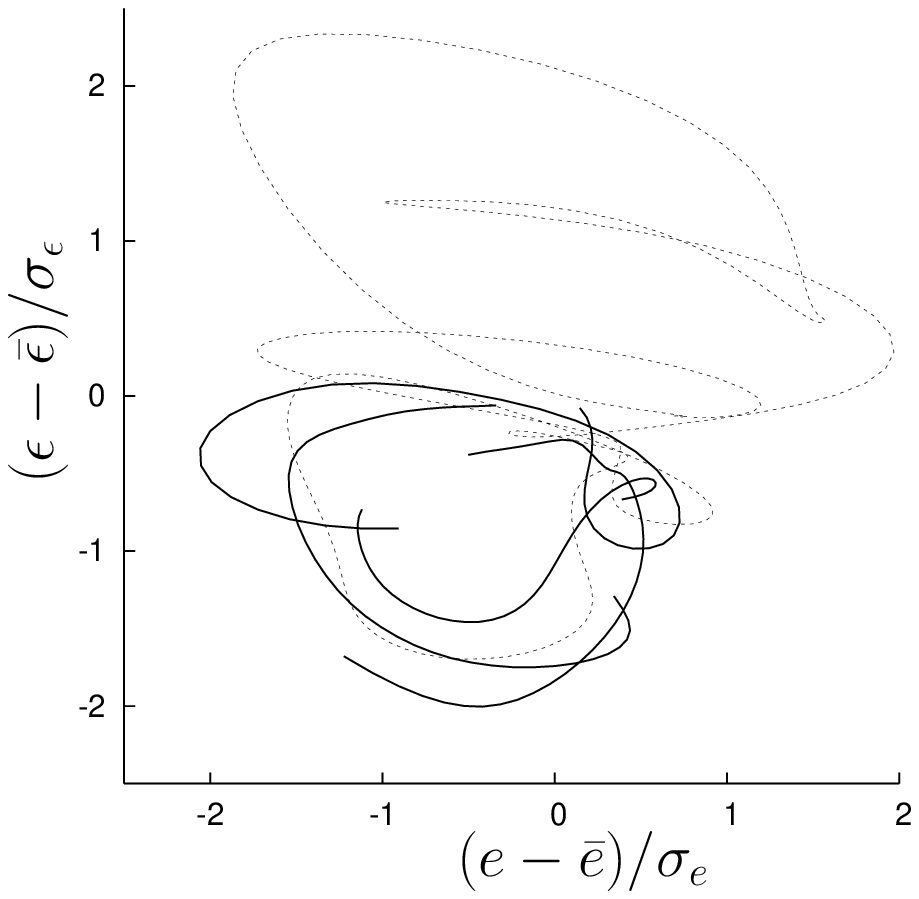}
\includegraphics[width=0.48\textwidth]{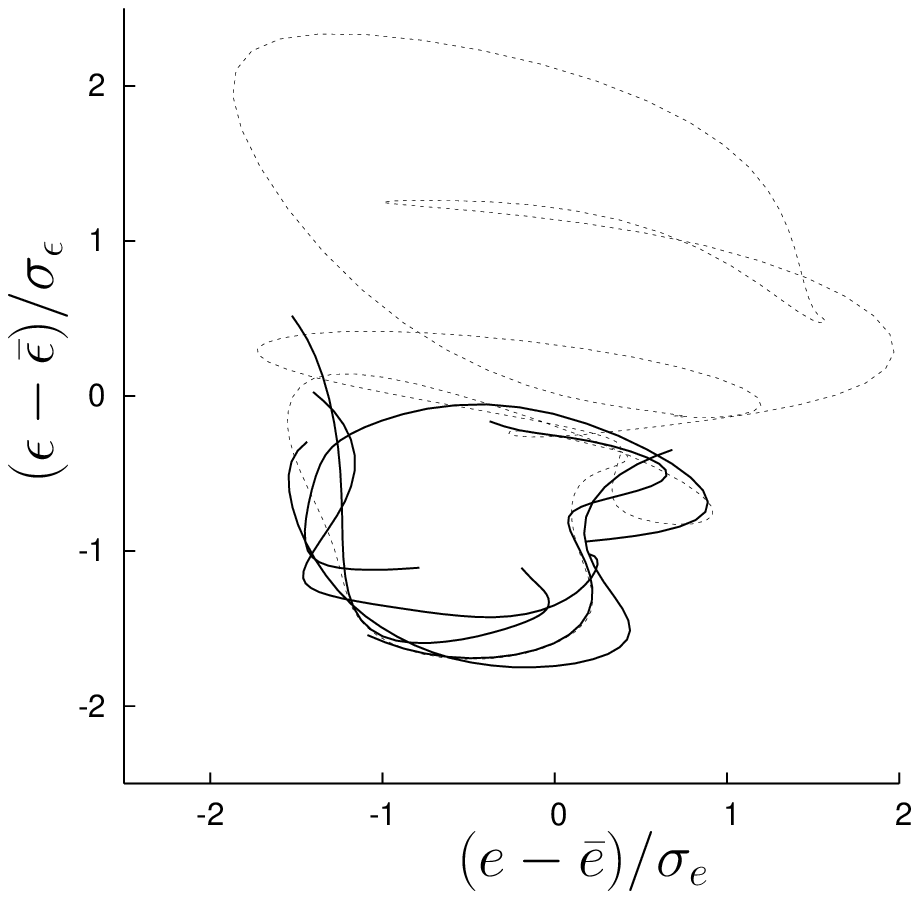}
\end{center}
\scaption{Turbulent orbits close to the period-5 motion. 
Such segments of length $T_{\rm{\scriptscriptstyle T}}$ that satisfy 
$D(t)<\overline{D}-1.5\sigma_{\rm{\scriptscriptstyle D}}$ 
in the $(e, \epsilon)$ plane are selected 
arbitrarily from a long turbulent orbit, and four examples are 
drawn in the respective figures. 
The dotted closed line indicates the priod-5 orbit. 
}
\label{fig:close}
\end{figure}

\hspace*{0.3cm} 
In order to get an idea of what the turbulent orbit looks like 
when it is approaching the period-5 motion, we show in 
Fig. \ref{fig:close} such segments of 
length $T_{\rm{\scriptscriptstyle T}}$ that satisfy 
$D(t)<\overline{D}-1.5\sigma_{\rm{\scriptscriptstyle D}}$ 
in the $(e, \epsilon)$ plane which are selected 
arbitrarily from a long turbulent orbit. 
Observe the way that the turbulent state is attracted around the period-5 
motion quite well, though such beautiful examples are not so frequent, 
i.e. only at the rate of once every $24T_{\rm{\scriptscriptstyle T}}$.

\vspace*{0.5cm}

\subsection{Lyapunov characteristics}
\label{subsection:lyapunov}

\hspace*{0.3cm}
Lyapunov exponents describe the growth or decay of perturbations with 
respect to a 
given reference solution of a dynamical system. The Lyapunov exponents
are a benchmark of chaos theory. If at least one exponent is positive, 
corresponding to a growing perturbation, the system is chaotic and there is 
sensitive dependence on initial conditions and a strange attractor with a 
fractal dimension. 
The rate at which information about the initial condition is lost and the 
dimension
of the chaotic attractor can be computed from the Lyapunov exponents.
Although turbulence can be regarded as a form
of high dimensional chaos, its Lyapunov characteristics are far from 
understood. 
Here, we will investigate the Lyapunov characteristics of isotropic 
turbulence by means of the embedded periodic solution.

\hspace*{0.3cm}
The Navier-Stokes equation (\ref{NS}) can be written symbolically as
\begin{equation}
\frac{\rm d}{{\rm d}t}{\bm{x}}=\bm{f}(\bm{x},\nu), \label{symb}
\end{equation}
where $\bm{x}\in \mathbb{R}^n$ is a vector that holds the Fourier transform of the
vorticity, $\widetilde{\bm{\omega}}$, and $\bm{f}$
denotes the right-hand side of Eq. (\ref{NS}). 
The linearised equations are then given by
\begin{equation}
\frac{\rm d}{{\rm d}t}{\bm{v}}=\bm{J}\bm{v}, \label{symb2}
\end{equation}
where $\bm{v}$ denotes a perturbation vorticity field 
$\delta\widetilde{\bm{\omega}}$, and $\bm{J}$
is the Jacobian matrix, i.e. 
$J_{ij}=\partial f_{i}(\bm{x}(t),\nu)/\partial x_j$.
The average rate of growth or decay of a perturbation is measured by 
the Lyapunov exponent
\begin{equation}
\itLambda = \lim_{t\rightarrow \infty} 
\frac{1}{2t}\ln \frac{\| \bm{v}(t)\|_{\rm{\scriptscriptstyle{Q}}}}{\|\bm{v}(0)\|_{\rm{\scriptscriptstyle{Q}}}}, 
\label{defLambda}
\end{equation}
where $\|\cdot\|_{\rm{\scriptscriptstyle{Q}}}$ again denotes the enstrophy norm
and a factor of $1/2$ is included because this norm is quadratic.

\hspace*{0.3cm}
In general, the value of $\itLambda$ depends on the reference solution 
$\bm{x}(t)$ and on
the initial perturbation $\bm{v}(0)$. However, in systems with a chaotic 
attractor the
Oseledec theorem guarantees that there is a spectrum of
limit values,
$\{\itLambda_{i}\}_{i=1}^{n}$, unique for the attractor.
At {\em almost every} initial point $\bm{x}(0)$
there are $n$ initial perturbations $\bm{v}_{i}(0)$ such that 
$\itLambda_i = \lim_{t\rightarrow \infty} (1/2t)\ln \left(
\| \bm{v}_i(t)\|_{\rm{\scriptscriptstyle{Q}}}/
\|\bm{v}_i(0)\|_{\rm{\scriptscriptstyle{Q}}}\right)$. 
The vectors
$\bm{v}_{i}(t)$ 
are called the Lyapunov vectors and depend on the initial point in a 
complicated manner. The Oseledec theorem holds for {\em almost every} initial point in the
basin of attraction of the chaotic attractor in a measure
theoretic sense. This means that starting from any generic initial condition we will find
the same Lyapunov spectrum, but for certain special initial points the spectrum may differ.
Examples of such special initial points are points lying on periodic solutions.

\hspace*{0.3cm}
The Lyapunov spectrum of chaotic motion can be used to measure the 
`strength' of the chaos or the complexity of the motion. 
Suppose that the Lyapunov exponents are ordered
such that $\itLambda_{1}>\itLambda_{2}>\ldots>\itLambda_{n}$, then the 
Kolmogorov-Sinai entropy
is defined by
\begin{equation}
H_{\rm{\scriptscriptstyle KS}}
=\sum_{i=1}^{k}\itLambda_{i}, \qquad \text{where}\ \ \itLambda_{k}>0 
\ \text{and} \ \ \itLambda_{k+1}<0 ,
\label{KSentropy}
\end{equation}
i.e. the sum of positive Lyapunov exponents, and the 
Kaplan-Yorke dimension 
is defined by 
\begin{equation}
D_{\rm{\scriptscriptstyle KY}}
=k +\frac{1}{|\itLambda_{k+1}|}\sum_{i=1}^{k}\itLambda_{i},
\qquad \text{where}\ \ \sum_{i=1}^{k}\itLambda_{i}>0 \ \text{and} \ \ 
\sum_{i=1}^{k+1}\itLambda_{i}<0 .
\label{KYdimension}
\end{equation}
One way to interpret these definitions is to realise that a volume element contained in the 
subspace spanned by any number of Lyapunov vectors will grow or decay at 
a rate given by the
sum of the corresponding Lyapunov exponents. Thus, the Kolmogorov-Sinai
entropy is the maximal rate of expansion for any volume element. It quantifies
the unpredictability of the dynamics. 
The Kaplan-Yorke 
dimension can be thought of
as the dimension of the chaotic attractor. Its integer part 
is the dimension of the largest volume element that
will grow in time, and the fractional part is added to render the function 
continuous in the Lyapunov exponents. 

\hspace*{0.3cm}
The numerical computation of more than only the leading Lyapunov exponent is troublesome
in systems with many degrees of freedom. The algorithms at hand require simultaneous
integration of several perturbation vectors and the repeated application of Gramm-Schmidt
orthogonalisation (see e.g. \citet{wolf}). This introduces numerical error, especially
when applied to truncations of the Navier-Stokes equation with small amplitude fluctuations
in the large wavenumber components. The limit in Eq.(\ref{defLambda}) has to 
be replaced
by an average over a finite time interval of the growth rate, and the convergence of
$\itLambda$ with time can be rather slow, of order $\mbox{O}(1/\sqrt{t})$. Consequently,
early attempts to compute a part of the Lyapunov spectrum for turbulent flows were
restricted to simulations at low resolution. Results for isotropic
turbulence \cite{grappin} and shear turbulence \cite{keefe} indicate that the
Kaplan-Yorke dimension is at least of order $\mbox{O}(100)$ even at low 
Reynolds number.

\hspace*{0.3cm}
In simulations at high resolution, the computation of a few hundred Lyapunov exponents
is hard if not impossible. One way around this problem is to inspect the {\em local} rather
than the {\em time average} growth rates. The local Lyapunov exponent can be defined by
\begin{equation}
\lambda(t)=\frac{1}{2}\frac{\mbox{d}}{\mbox{d}t}\ln \|\bm{v}(t)\|_{\rm{\scriptscriptstyle{Q}}} ,
\label{deflocal}
\end{equation}
such that, taking the time average, we have 
$\overline{\lambda}=\itLambda$. 
The evolution
of the Lyapunov vectors and the associated local Lyapunov exponents was studied in 
the case of weakly turbulent Taylor-Couette flow by
\citet{vastano}. They managed to tie the
local Lyapunov exponents and vectors to physical instabilities in the transition to
chaotic behaviour. 

\hspace*{0.3cm}
In the same spirit we seek to investigate the Lyapunov characteristics of developed
isotropic turbulence. For this purpose we use the period-5 orbit as the reference
solution. The choice of a periodic reference solution greatly facilitates the analysis.
Let $\bm{x}(t)$ be a solution of Eq.(\ref{symb}) such that $\bm{x}(t+T)=\bm{x}(t)$
for all $t$ and some period $T$. By Floquet theory the solution of Eq.(\ref{symb2}) 
can then be written as
\begin{equation}
\bm{v}(t)=\bm{M}(t)\mbox{e}^{\bm{\scriptstyle A}t}\bm{v}(0),
\label{Floquet}
\end{equation}
where $\bm{M}(t)=\bm{M}(t+T)$ is a periodic matrix satisfying 
$\bm{M}(0)=\mathbb{I}$ (unit matrix), and 
$\bm{A}$ is a constant matrix.
Thus we find that the Lyapunov spectrum $\{\itLambda_i,\bm{v}_{i}(t)\}$ is 
determined by
the eigenspectrum $\{\mu_i,\bm{w}_i\}$ of $\bm{A}$. 
For each real eigenvalue $\mu_i$
we have $\itLambda_i=\mu_i$, $\bm{v}_i(0)=\bm{w}_i$, and for the local exponent we find
\begin{equation}
\lambda_i (t)=\itLambda_i + \frac{1}{2}\frac{\mbox{d}}{\mbox{d}t}\ln 
\|\bm{M}(t)\bm{v}_i(0)\|_{\rm{\scriptscriptstyle{Q}}}.
\label{locallyap}
\end{equation}
For each complex pair $\{\mu_i,\mu_{i+1}\}$ 
we have $\itLambda_i=\itLambda_{i+1}=\text{Re}\left(\mu_i\right)$,
$\bm{v}_i(0)=\text{Re}\left(\bm{w}_i\right)$, 
$\bm{v}_{i+1}(0)=\text{Im}\left(\bm{w}_i\right)$ and
\begin{equation}
\lambda_i (t)=\lambda_{i+1}(t)=\itLambda_i + 
\frac{1}{2}\frac{\mbox{d}}{\mbox{d}t}\ln \left(
\|\bm{M}(t)\bm{v}_i(0)\|_{\rm{\scriptscriptstyle{Q}}}
+\|\bm{M}(t)\bm{v}_{i+1}(0)\|_{\rm{\scriptscriptstyle{Q}}} \right).
\end{equation}

\hspace*{0.3cm}
The matrix $\mbox{e}^{\bm{\scriptstyle A}T}$ is computed in much the same way 
as we computed the matrix
of derivatives of the Poincar\'e map as described in section \ref{sec:periodic}. We then
solve the eigenvalue problem to find any number of Lyapunov exponents and vectors.
In order to find the $\lambda_i (t)$ we integrate the linearised Navier-Stokes equations
along the period-5 orbit with the eigenvectors $\bm{v}_{i}(0)$ as initial 
condition.
In this integration numerical errors in $\bm{v}_{i}(t)$ tend to grow as 
$\exp([\itLambda_1-\itLambda_i]t)$, which puts a limit to the number of local
Lyapunov exponents we can compute. The results presented below are based on analysis
of the first $50$ exponents. The largest and smallest average exponents are 
$\itLambda_1=0.238$ and $\itLambda_{50}=-0.584$.

\begin{figure}[t]
\begin{center}
\includegraphics[width=0.7\textwidth]{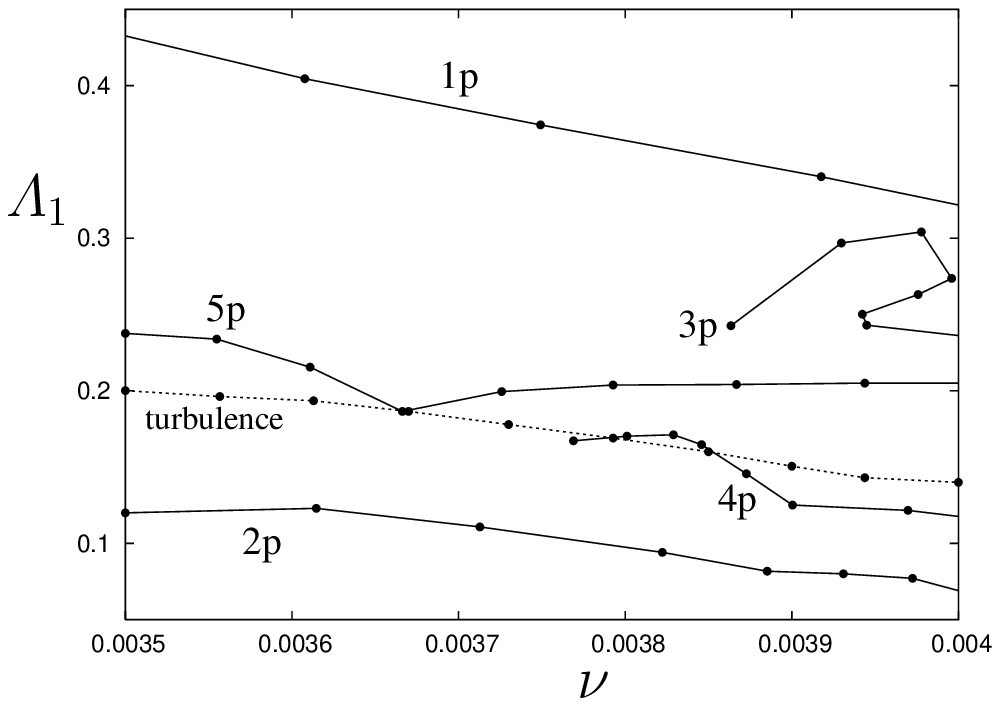}
\end{center}
\scaption{The largest Lyapunov exponent $\itLambda_1$ measured along the 
periodic orbits, labeled as in figure \ref{continuation}. The value found for turbulent
motion is represented by the dotted line.
}
\label{LLEnu}
\end{figure}

\hspace*{0.3cm}
As mentioned above, the Lyapunov spectrum of periodic motion is different 
from that of turbulent motion. However, as argued in the preceding subsections
we consider the periodic motion as the skeleton of turbulence, and its 
qualitative properties as an approximation of the corresponding properties of turbulence.
Thus the Lyapunov characteristics of the periodic motion are expected to be close to those
of turbulence.
A direct comparison is given by the leading Lyapunov exponent,
which can be computed for the turbulent motion as described in \citet{kida3}. Fig. \ref{LLEnu}
shows $\itLambda_1$ for the turbulent motion and for the five periodic orbits in the
parameter range $0.0035<\nu<0.004$. As we saw when comparing the energy-dissipation rate of periodic and 
turbulent motion in Fig. \ref{continuation}, the period-5 orbit reproduces the values
found for turbulent motion well, whereas the shorter periodic orbits deviate. 
At the time of writing, the continuation curves for the period-3 and period-4 orbits were
incomplete. Further computations are in progress.
The numerical values at $\nu=0.0035$ are $\itLambda_{1}=0.2$ and $\itLambda_{1}^{5\rm{p}}=0.238$
for the turbulent and the period-5 motion, respectively.

\begin{figure}[t]
\begin{center}
\includegraphics[width=0.7\textwidth]{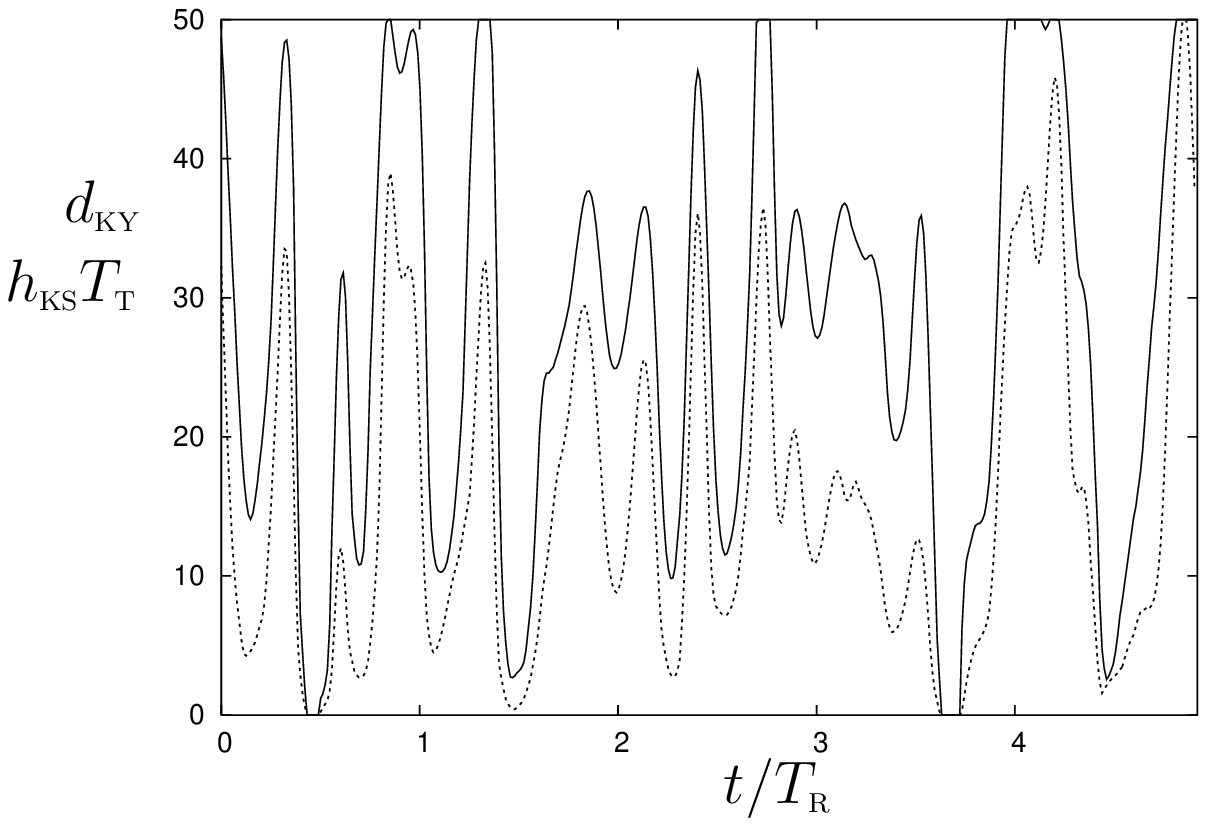}
\end{center}
\scaption{The local Kaplan-Yorke dimension 
$d_{\rm{\scriptscriptstyle KY}}(t)$ (solid line) and the local 
Kolmogorov-Sinai entropy $h_{\rm{\scriptscriptstyle KS}}(t)$ 
(nondimensionalised by $T_{\rm{\scriptscriptstyle T}}$ and drawn with a dotted line) as computed
from the leading 50 local Lyapunov exponents $\{\lambda_{i}(t)\}_{i=1}^{50}$ 
along the period-5 orbit. Note the
rapid oscillations and large amplitude.
}
\label{KYSK}
\end{figure}

\hspace*{0.3cm}
As we know only the leading Lyapunov exponent for turbulent flow, we cannot directly
estimate $H_{\rm{\scriptscriptstyle KS}}$ and $D_{\rm{\scriptscriptstyle KY}}$ 
for the chaotic attractor. For the period-5 motion we find that 
$H_{\rm{\scriptscriptstyle KS}}^{5\rm{p}}=0.992$ and 
$D_{\rm{\scriptscriptstyle KY}}^{5\rm{p}}=19.7$. Note, that these values 
cannot directly be compared to those for general isotropic turbulence because 
we can only compute the
contribution of perturbations that satisfy the high-symmetry constraints 
described in section \ref{sec:highsymm}. 
In the full phase space, without any symmetry constraints,
these values are likely to be a factor of order $\mbox{O}(100)$ times higher.
The {\it local} Kolmogorov-Sinai entropy 
$h_{\rm{\scriptscriptstyle KS}}(t)$ 
and {\it local} Kaplan-Yorke dimension 
$d_{\rm{\scriptscriptstyle KY}}(t)$
can be computed from the local Lyapunov exponents, substituting 
the $\lambda_i(t)$ 
for the $\itLambda_i$ in Eqs. (\ref{KSentropy}) and (\ref{KYdimension}). 
Thus, we 
get an impression of the change of the complexity of the flow with time.
The graph is shown in Fig. \ref{KYSK}. Note that, strictly speaking, 
we can only
compute a lower bound for the local quantities as we only know the leading
50 local Lyapunov exponents. However, $\lambda_{50}(t)$ is negative at all 
times and we expect only minor contributions, if any, from higher exponents.

\hspace*{0.3cm}
The local Lyapunov exponents and the derived
quantities $h_{\rm{\scriptscriptstyle KS}}(t)$ and 
$d_{\rm{\scriptscriptstyle KY}}(t)$
show large fluctuations on a time scale as short as the Kolmogorov dissipation
time scale $\tau_{\eta}=\sqrt{\nu/\bar{\epsilon}}\approx 0.2$. 
Around 
$t/T_{\rm{\scriptscriptstyle R}}=4$
in the active
phase identified in section \ref{subsec:structure}, 
$d_{\rm{\scriptscriptstyle KY}}(t)$
jumps from $0$ to a wide maximum larger than $50$ and back. 
This peak coincides with
the dominant peak of the energy-dissipation rate. 
The second wide maximum lies around
$t/T_{\rm{\scriptscriptstyle R}}=4.9$
and coincides with a peak of the energy-input rate. 
Thus it seems that the local Lyapunov 
exponents and the complexity of the flow are correlated to physical, 
spatial mean quantities.

\begin{figure}[t]
\begin{center}
\includegraphics[width=0.9\textwidth]{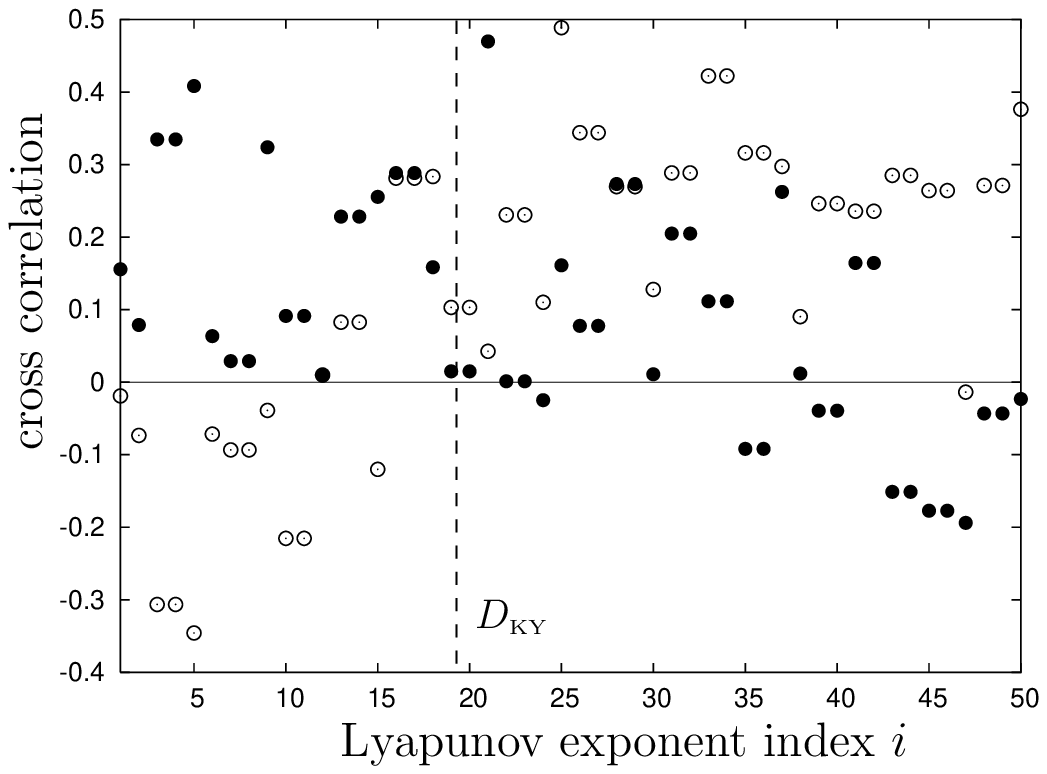}
\end{center}
\scaption{Coefficient of correlation of the local Lyapunov exponents 
$\lambda_{i}(t)$ with the energy-input rate (filled circles) and 
the energy-dissipation rate (open circles). Despite a fair amount of
scatter we can see that the exponents 
$\lambda_{i}(t)$ with $1\leq i <D_{\rm{\scriptscriptstyle KY}}$
behave differently from those of higher indices.
}
\label{correlation}
\end{figure}

\hspace*{0.3cm}
In order to check this conjecture we compute the correlation between the 
$\lambda_{i}(t)$ on one hand, and $e(t)$ and $\epsilon(t)$ on the other. 
The correlation coefficients are defined by
\begin{eqnarray}
c^{i}_{e}=\frac{1}{T_{5\rm{p}}\sigma_{e}^{5\rm{p}}\sigma_{\lambda_i}}
\int_{0}^{T_{5\rm{p}}}(e^{5\rm{p}}(t)-\bar{e}^{5\rm{p}})
(\lambda_i(t)-\itLambda_i) \mbox{d}t, \nonumber \\[0.5cm]
c^{i}_{\epsilon}=\frac{1}{T_{5\rm{p}}
\sigma_{\epsilon}^{5\rm{p}}\sigma_{\lambda_i}}
\int_{0}^{T_{5\rm{p}}}(\epsilon^{5\rm{p}}(t)-\bar{\epsilon}^{5\rm{p}})
(\lambda_i(t)-\itLambda_i) \mbox{d}t,
\end{eqnarray}
where $\sigma_{\lambda_i}$ is the standard deviation of $\lambda_i(t)$. 
Fig. \ref{correlation}
shows $c^{i}_{e}$ and $c^{i}_{\epsilon}$ for the first 50 local Lyapunov 
exponents.
Although there is a lot of scatter in the data, a structural difference 
between the
local Lyapunov exponents with a small and a large index is obvious. 
Those with a small
index have a negative correlation with the energy-dissipation rate and a 
positive correlation with the energy-input rate,
whereas those with a large index have a positive correlation with 
the energy-dissipation rate and a correlation
of either sign with the energy-input rate. 
This suggests that the Lyapunov vectors have a preferred spatial scale. 
In particular, we conjecture that the Lyapunov vectors with a small index
describe perturbation fields with a large spatial scale, 
directly excited by the energy input. 
Those with a larger index describe smaller scale perturbation fields and 
are more strongly correlated to energy dissipation.

\begin{figure}[t]
\begin{center}
(a)
\includegraphics[width=0.7\textwidth]{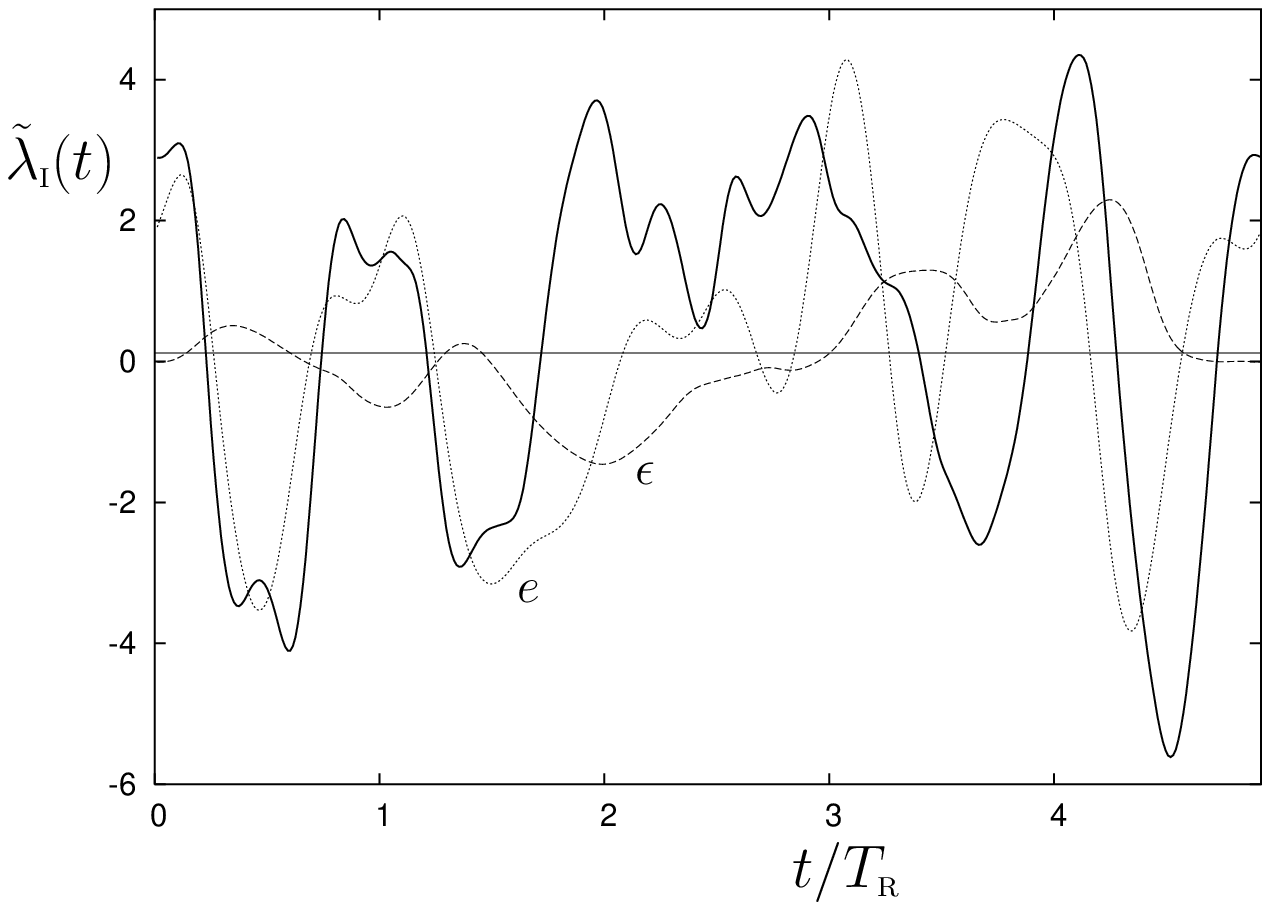}

\vspace{0.5cm}

(b)
\includegraphics[width=0.7\textwidth]{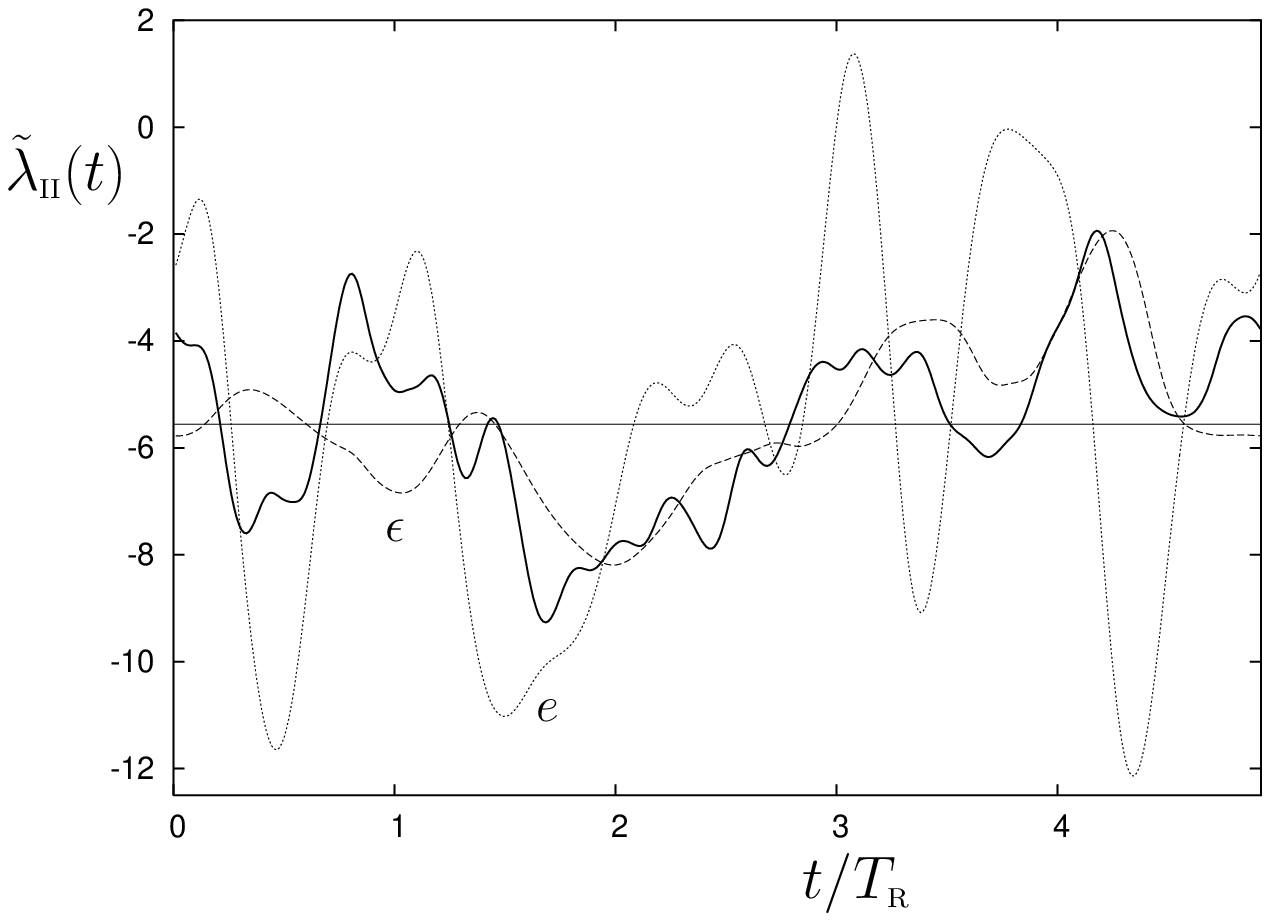}
\end{center}
\scaption{
Temporal variation of (a) $\widetilde{\lambda}_{\rm{\scriptscriptstyle I}}(t)$ 
and (b) $\widetilde{\lambda}_{\rm{\scriptscriptstyle II}}(t)$. 
The running average is taken over 
$\tau_{\rm{av}}
= 0.54 T_{\rm{\scriptscriptstyle R}}$.
For comparison the energy-input and dissipation rates, 
shifted and scale by equal amounts, are drawn with a
dotted and a dashed line, respectively. 
The horizontal lines indicate the time mean values, 
$\overline{\lambda}_{\rm{\scriptscriptstyle I}}=0.122$ and 
$\overline{\lambda}_{\rm{\scriptscriptstyle II}}=-5.57$.
}
\label{fig:local-lyapunov}
\end{figure}

\hspace*{0.3cm}
In order to test this conjecture we divide the Lyapunov spectrum into two 
parts with an equal number of exponents. 
As indicated in Fig. \ref{correlation}, we choose 
the integer part of the Kaplan-Yorke dimension, computed from the time 
averaged Lyapunov 
exponents, to separate the two. 
Thus, group I comprises $\{\itLambda_i,\bm{v}_i\}_{i=1}^{19}$
and group II comprises $\{\itLambda_i,\bm{v}_i\}_{i=20}^{38}$. 
The growth rate of
volumes in these two subspaces is given by
$\lambda_{\rm{\scriptscriptstyle I}}(t)=\sum_{i=1}^{19}\lambda_{i}(t)$ and
$\lambda_{\rm{\scriptscriptstyle II}}(t)=\sum_{i=20}^{38}\lambda_{i}(t)$, 
respectively.
As mentioned above, the $\lambda_{i}(t)$
fluctuates rapidly. 
In order to see a possible correlation with the energy-input and 
dissipation rates 
we compute the running mean of $\lambda_{\rm{\scriptscriptstyle I}}(t)$ and
$\lambda_{\rm{\scriptscriptstyle II}}(t)$ over a time interval 
$\tau_{\rm{av}}$ such that
$\tau_{\eta}<\tau_{\rm{av}}<T_{\rm{\scriptscriptstyle R}}$. 
The running mean is indicated by a tilde.
Figs. \ref{fig:local-lyapunov}(a) and (b) show the time series of 
$\widetilde{\lambda}_{\rm{\scriptscriptstyle I}}(t)$ 
and $\widetilde{\lambda}_{\rm{\scriptscriptstyle II}}(t)$ along with 
the energy-input and dissipation rates, shifted to have
the same time mean value and scaled by equal factors. Clearly, 
$\widetilde{\lambda}_{\rm{\scriptscriptstyle I}}(t)$ has a strong 
positive correlation with the energy-input rate 
and a weaker, negative correlation with the energy-dissipation rate. 
Most of the peaks of 
$\widetilde{\lambda}_{\rm{\scriptscriptstyle I}}(t)$ coincide with peaks of 
the energy-input rate, the latter leading
in phase. Only in the interval $2<t/T_{\rm{\scriptscriptstyle R}}<3$ the correlation is
not very clear. 
In contrast, $\widetilde{\lambda}_{\rm{\scriptscriptstyle II}}(t)$ shows a 
strong 
positive correlation with the energy-dissipation rate and tends to lead 
in phase. On the interval 
$0<t/T_{\rm{\scriptscriptstyle R}}<1$ the correlation with the energy-dissipation rate is weaker, and locally
there is a positive correlation with the energy-input rate.

\begin{figure}[t]
\begin{center}
\includegraphics[width=0.7\textwidth]{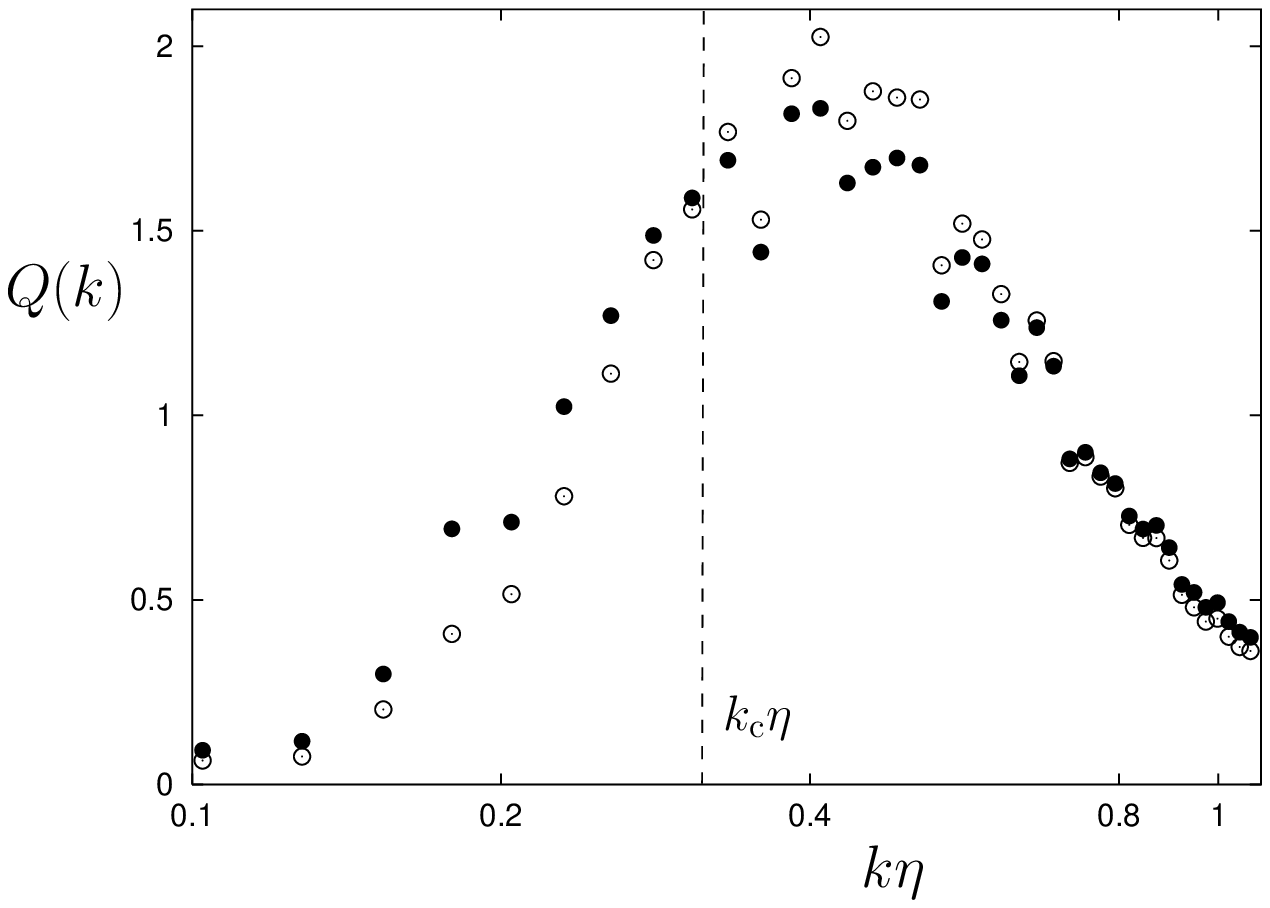}
\end{center}
\scaption{Enstrophy spectra $Q_{\rm{\scriptscriptstyle I}}(k)$ and $Q_{\rm{\scriptscriptstyle II}}(k)$
of the perturbation vorticity fields $\delta\hat{\bm{\omega}}_{i}$ in the groups I and II.
The filled circles represent
the average profile of Lyapunov vectors 1 through 19, the open circles
represent the average profile of Lyapunov vectors 20 through 38. The average profile of
the leading 19 Lyapunov vectors is larger for $k\eta < k_{\rm{c}}\eta\approx 0.32$ and 
mostly smaller for larger wave numbers.}
\label{support}
\end{figure}

\hspace*{0.3cm}
Finally we consider the orientation of the Lyapunov vectors. 
Consider the Lyapunov
vectors scaled to unit length, 
$\hat{\bm{v}}_i=\bm{v}_i(t)/\|\bm{v}_i(t)\|_{\rm{\scriptscriptstyle{Q}}}^{1/2}$ and denote the corresponding
perturbation vorticity field by $\delta\hat{\widetilde{\bm{\omega}}}_i$. 
We compute the 
enstrophy spectrum of the scaled perturbation fields as
\begin{equation}
Q_i(k)=\frac{1}{2}\sum_{k-\frac{1}{2}
<\|\bm{k}\|<k+\frac{1}{2}}|\delta\hat{\widetilde{\bm{\omega}}}_i
(\bm{k})|^2.
\end{equation}
If the Lyapunov vectors have a preferred length scale we expect to see a structural difference
between the enstrophy spectra of the vectors in group I and group II, the former having a larger
amplitude in the smaller wavenumbers and the latter in the larger wavenumbers. 
In Fig. \ref{support}, 
we plot the average spectrum over the perturbation fields in group I, 
$Q_{\rm{\scriptscriptstyle I}}(k)$,
and group II, $Q_{\rm{\scriptscriptstyle II}}(k)$. 
All perturbation
fields have the maximal amplitude around $k\eta=0.4$, corresponding to a 
spatial scale in between
that of the fixed modes, $k_{f}^{-1}$, and the Kolmogorov dissipation scale 
$\eta$ in the current simulations.
The average
spectrum $Q_{\rm{\scriptscriptstyle I}}(k)$ is larger for all wavenumbers 
below $k_{\rm{c}}$ ($\approx 0.32$) and smaller
for most larger wavenumbers.

\hspace*{0.3cm}
These results indicate that the Lyapunov vectors indeed have preferred length scales associated
with them, and that the local Lyapunov exponents are correlated with 
the physical 
quantities that dominate these spatial scales. We have checked that the results do not depend
critically on the choice of the two groups, i.e. the highest index in group I, here fixed
to the integer part of the Kaplan-Yorke dimension $D_{\rm{\scriptscriptstyle KY}}$.
As far as we know, 
this is the first time that evidence is found for the localisation of
Lyapunov vectors and the correlation of (local) Lyapunov exponents and physical quantities
in developed turbulence.
The localisation of Lyapunov vectors has been found in shell model turbulence by \citet{yamada}
(and references therein).
However, their results are derived at much larger Reynolds number, in the presence of a
large inertial range. We expect that the presence of a developed inertial range in the
case of isotropic turbulence would yield an even clearer separation of spatial scales than
seen in our present results.

\section{Concluding Remarks}
\label{sec:conclusion}

\hspace*{0.3cm}
We have identified temporally periodic motion which reproduces 
the dynamics and statistics of isotropic turbulence well  
in high-symmetric flow. 
The period of the periodic motion is of the order of the
eddy-turnover time of turbulence. 
The mean properties of various physical quantities, 
e.g. the energy spectral function and the Lyapunov exponent,  
calculated by time average taken over one period of the periodic motion 
approximate those of the turbulence taken over a long time series. 
This agreement may be understood by noting the fact that the turbulent motion 
spends much of the time in the same, or similar, 
spatio-temporal state as the periodic motion. 
In fact, we have seen
that the state point of the turbulent motion 
approaches the orbit of the periodic motion in phase space at the rate of once 
over several eddy-turnover times. 
In other words, the orbit of this periodic motion is 
embedded in turbulence. 
Thus, we regard it as the skeleton of turbulence. 

\hspace*{0.3cm}
Such a periodic motion embedded in turbulence 
is useful as a reference field with respect to 
which the mechanisms of various turbulence phenomena, including 
turbulent mixing and the energy-cascade process, can be analysed. 
The reason is as follows. 
Turbulence is intrinsically chaotic and the fluid flow varies 
quite randomly both in space and in time. 
The fluid motion is unpredictable and never repeats, though 
the statistical properties are rather universal. 
This chaotic nature makes it difficult to study the general 
properties of turbulence. 
There is no way to confirm that those turbulence data used in 
analysis represent typical properties of turbulence. 
On the other hand, the periodic motion, whose dynamical properties 
can be understood much more clearly than those of the turbulence itself, 
repeats exactly its temporal variation without limit. 
Then, by analysing the repeated, periodic time series 
we may be able to extract the typical mechanisms of turbulence 
dynamics as well as calculate the statistics of any physical 
quantities with high accuracy. 
This line of study is now under way. 

\hspace*{0.3cm}
In the present study the inertial range is captured only marginally. 
In order to increase the resolution it is necessary to simulate high-Reynolds 
number turbulence. 
The difficulties then arise in the computation time and memory requirements
of the algorithm used to find periodic motion. 
The calculation of iteration matrix of the 
Newton-Raphson procedure is most time-consuming. 
The number of degrees of freedom in our computations, $\rm{O}(10^4)$, 
seems to be the maximal number tackled in continuation of periodic orbits
at the time of writing. 
It proved possible to use the conventional 
arc-length method with the Newton-Raphson iteration because
of the high efficiency of parallelization. 
In order to go to higher truncation levels,
and thus larger Reynolds numbers and a developed inertial range, it may
be necessary to switch to matrix-free methods for continuation. 
Such methods use inexact linear solvers for equations like (\ref{fixed}), 
avoiding the computation 
and orthogonal decomposition of the matrix of derivatives. 
Recently this approach
has successfully been applied to the computation of periodic solutions of the
Navier-Stokes equations \cite{sanchez}.

\hspace*{0.3cm}
It is conjectured that there are infinitely many periodic orbits in 
the turbulent regime. 
Some of them represent turbulent state and the others do not. 
From the present study we cannot infer a general rule for selecting
periodic motion which represents the turbulence well. 
It is interesting
to see, that at low micro-scale Reynolds number, where we distill the periodic
orbits from a turbulent time series, their time mean energy-dissipation rate
and largest Lyapunov exonent are all close to those of turbulence. 
If we decrease the viscosity, however, only the orbit of longest period 
reproduces the average values of turbulence. 
Future work should aim at a better understanding of this selection process 
as well as of the uniqueness of such solutions. 

\section*{Acknowledgments}

Author L. van Veen was supported by a grant of the Japan Society for Promotion 
of Science. The parallel computations were done at the Media Center of Kyoto
University.

\end{document}